\documentclass[12pt,a4paper]{article}
\usepackage{amsmath}
 \usepackage{bm}
 \usepackage{natbib}
  \usepackage{setspace}
  \usepackage{xcolor}
   \usepackage{graphicx}
  \usepackage{subfigure}
  \usepackage{algorithm}
  \usepackage{algpseudocode}
  \usepackage{multirow}
  \usepackage{mathtools}  
  \usepackage{url}
  \usepackage{authblk}
  
\algblock{ParFor}{EndParFor}
\algnewcommand\algorithmicparfor{\textbf{parfor}}
\algnewcommand\algorithmicpardo{\textbf{do}}
\algnewcommand\algorithmicendparfor{\textbf{end\ parfor}}
\algrenewtext{ParFor}[1]{\algorithmicparfor\ #1\ \algorithmicpardo}
\algrenewtext{EndParFor}{\algorithmicendparfor}

\newcommand{\Yb}{\bm{Y}}

\newcommand{\ib}{\bm{i}}

\newcommand{\thetab}{\bm{\theta}}
\newcommand{\thb}{\bm{\theta}}
\newcommand{\ths}{\theta^\star}
\newcommand{\thbs}{\bm{\theta^\star}}

\newcommand{\xib}{\bm{A}}                 %% changed
\newcommand{\pixi}{\pi_{\Ab}}

\newcommand{\Pib}{\bm{F}}                 %% changed
\newcommand{\Fb}{\bm{F}}
\newcommand{\etab}{\bm{\eta}}
\newcommand{\pib}{\bm{\pi}}

\newcommand{\Cb}{\bm{C}}
\newcommand{\Bb}{\bm{B}}

\newcommand{\yb}{\bm{y}}

\newcommand{\Ab}{\bm{A}}

\newcommand{\mub}{\bm{\mu}}

\newcommand{\Sigmab}{\bm{\Sigma}}

\newcommand{\ybt}{\widetilde{\yb}}

\renewcommand{\th}{\theta}
\newcommand{\eps}{\epsilon}

\newcommand{\DP}{\mbox{DP}}
\newcommand{\IBP}{\mbox{IBP}}
\newcommand{\Bin}{\mbox{Bin}}

\newcommand{\Poi}{\mbox{Poi}}

\newcommand{\ind}{\stackrel{\mbox{\scriptsize ind}}{\sim}}
\newcommand{\iid}{\stackrel{\mbox{\scriptsize iid}}{\sim}}

\newcommand{\bch}{\color{black}\rm}

\newcommand{\ech}{\color{black}\rm}

\title{\bf Consensus Monte Carlo for Random Subsets using Shared Anchors}

\author[1]{Yang Ni}
\author[2]{Yuan Ji}
\author[3]{Peter M\"uller}
\affil[1]{Department of Statistics, Texas A\&M University}
\affil[2]{Department of Public Health Sciences, The University of Chicago}
\affil[3]{Department of Mathematics, The University of Texas at Austin}

\date{}

\begin{document}
\def\spacingset#1{\renewcommand{\baselinestretch}%
	{#1}\small\normalsize} \spacingset{1}
\maketitle

\bigskip

\begin{abstract}
We present a consensus Monte Carlo algorithm that scales existing
Bayesian nonparametric models for clustering and feature allocation to
big data.  The algorithm is valid for any prior on random
subsets such as partitions and latent feature allocation, under essentially any sampling model. 
Motivated by three case studies, we focus on clustering
induced by a Dirichlet process mixture sampling model, inference
under an Indian buffet process prior with a binomial sampling
model, and with a categorical sampling model. We assess  the proposed algorithm with
simulation studies and show results for inference with three datasets:
an MNIST image dataset, a dataset of pancreatic cancer
mutations, and a large set of electronic health records (EHR). Supplementary materials for this article are available online.
\end{abstract}

\noindent%
{\it Keywords:} Big data, electronic health records, image cluster, parallel computing, tumor heterogeneity, 
\vfill

\newpage
\spacingset{1.45} % DON'T change the spacing!
 \section{Introduction}
We develop a consensus Monte Carlo (CMC) algorithm for
Bayesian nonparametric (BNP) inference with large datasets that 
are too big for full posterior simulation on a single machine,
due to CPU or memory limitations.
The proposed algorithm  is for inference under BNP models for
random subsets, including clustering, feature allocation (FA),
and related models.
We distribute a large dataset to multiple machines, run separate  instances of Markov
chain Monte Carlo (MCMC) simulations in parallel and then aggregate the
Monte Carlo samples across machines. The idea of the proposed CMC
hinges on choosing a portion of observations as \textit{anchor points}
\citep{kunkel2018anchored} which are distributed to every machine
along with other observations that are only available to one
machine. Those anchor points then serve as anchors to merge Monte
Carlo draws of clusters or features across machines.

Clustering is
an unsupervised learning method that partitions observations into
non-overlapping subsets (clusters)  with the aim of creating
homogeneous groups. A widely used approaches for model-based inference on random
partitions is based on mixtures, with each mixture component being a
cluster. Bayesian finite mixture models with a random number of terms
that allow for inference with an {\em a priori} unknown number of clusters,
is first discussed in \cite{richardson1997bayesian}.
A natural generalization of such models are infinite mixtures
with respect to discrete random probability measures. In fact, any exchangeable random partition can be represented
this way \citep{Kingman78}. Prior probability models for random probability measures, such as the mixing measure in this representation, 
are known as BNP models.
Examples include Dirichlet process mixtures (DPM,
\citealt{lo1984class,maceachern2000dependent,lau2007bayesian}),
Pitman-Yor process mixtures \citep{pitman1997two} and variations with
different data structures, such as \cite{rodriguez2011sparse} for
mixtures of graphical models, normalized inverse Gaussian process
mixtures \citep{lijoi2005hierarchical}, normalized generalized Gamma
process mixtures \citep{lijoi2007controlling}, and more general
classes of BNP mixture models
\citep{barrios&al:13,favaro&teh:13,Argiento:2010}. For a more general discussion of BNP priors on random partitions see
also de Blasi et al. (2015). \nocite{deBlasiAl:14}

Feature allocation, also known as overlapping clustering, relaxes the
restriction to mutually exclusive and non-overlapping subsets and
allocates each observation to possibly more than one subset (``feature'').
The most commonly used feature allocation model is the Indian buffet process
(IBP,
\citealt{ghahramani2006infinite,broderick2013cluster,broderick2013feature}). 

A convenient way to represent random subsets, in clustering as well as
in feature allocation, is as a binary 
matrix $\Ab$ with $A_{ik}=1$ indicating that experimental unit $i$ is a
member in the $k$-th random subset. Feature allocation for $n$
experimental units into an unknown number $K$ of subsets then takes
the form of a prior $p(\Ab)$ for a random binary $(n \times K$)
matrix $\Ab$ with a random number of columns. 
Similarly, a random partition becomes a prior $p(\Ab)$ with exactly
one non-zero element in each row, and a random number $K$ of columns,
$1 \leq K \leq n$. 
An important feature of BNP clustering and feature
allocation models is that they do not require an {\em a prior}
specification of the number of subsets, (clusters or
features). An important limitation is the requirement for
intensive posterior simulation. Implementation of posterior
inference by MCMC simulation is usually not scalable to large
datasets. 

Several approaches have been proposed to overcome these
computational limitations in general big-data problems, not necessarily restricted to BNP models. 
\cite{huang2005sampling}
proposed CMC algorithms that distribute data to multiple machines,
each of which runs a separate MCMC simulation in parallel.  Various
ways of consolidating simulations from these local posteriors have
been proposed. \cite{scott2016bayes} combined the local posterior
draws by weighted averages. \cite{neiswanger2013asymptotically} and
\cite{white2015piecewise} proposed parametric, semiparametric and
nonparametric approaches to approximate the full posterior density as
the product of local posterior
densities. \cite{wang2013parallelizing} introduced the Weierstrass sampler
which applies a Weierstrass transform to each local
posterior. \cite{minsker2014scalable} found the geometric median of
local posterior distributions using a reproducing kernel Hilbert space
embedding.  \cite{rabinovich2015variational} proposed a
variational Bayes algorithm that optimizes over aggregation functions
to achieve a better approximation to the full posterior. \bch
Recently, \cite{entezari2018likelihood} introduced 
the likelihood inflating sampling algorithm (LISA) as 
a clever alternative to CMC. Instead of raising the prior to a
fractional power as in CMC, LISA  inflates the likelihood so that each
local posterior is a close representation of the whole-data
posterior.  \ech 
All but the method by \cite{rabinovich2015variational} are
specifically designed to aggregate global parameters. However, local
parameters i.e., parameters that are indexed by experimental units,
such as
cluster assignment and latent feature allocation are also of great
importance in big data analytics.  Even though the method by
\cite{rabinovich2015variational} can be used to aggregate local
structures such as cluster assignment, it assumes the number of
clusters is fixed a priori, which limits it applicability. Our
contribution is to bridge this gap in the literature.

\cite{blei2006variational}
developed the first variational Bayes algorithm for posterior
inference of DPM which was later extended to several variational
algorithms by \cite{kurihara2007collapsed}. A similar variational
Bayes algorithm for posterior inference under an IBP prior was
derived by
\cite{doshi2009variational}. \cite{doshi2009large} propose a
parallel MCMC algorithm which relies on efficient message passing
between the root and worker processors.  \cite{wang2011fast} developed
a single-pass 
sequential algorithm for conjugate DPM models. In each iteration, they
deterministically assign the next subject to the cluster with the
highest probability conditional on past cluster assignments and data
up to the current observation.  \cite{rai2011beam} developed a
beam-search heuristic to perform approximate maximum a posteriori
inference for linear-Gaussian models with IBP prior. Under the same
model, \cite{reed2013scaling} proposed a greedy
maximization-expectation algorithm, a special case of variational
Bayes algorithm, which exploits the submodularity of the evidence
lower bound.  \cite{broderick2013mad} and \cite{xu2015mad}  develop small-variance asymptotics for approximate posterior inference under an IBP prior.
\cite{Lin:13} proposed a one-pass sequential algorithm for DPM
models. The algorithm utilizes a constructive characterization of the
posterior distribution of the mixing distribution given data and a
partition. Variational inference is adopted to sequentially
approximate the marginalization. \cite{WilliamsonAl:13} introduced a
parallel MCMC for DPM models which involves iteration over local
updates and a global update. For the local update, they exploit the
fact that a Dirichlet mixtures of Dirichlet processes (DP) again
defines a DP, if the parameters of Dirichlet mixture are suitably
chosen.  \cite{GeAl:15} used a similar characterization of the DP as
in \cite{Lin:13}. But instead of a variational approximation, they
adapted the slice sampler for parallel computing under a MapReduce
framework. \cite{TankAl:15} developed two variational inference
algorithms for general BNP mixture models.  Recently,
\cite{zuanetti2018Bayesian} suggested two efficient alternatives using
DPM with conjugate priors. The first approach is based on a predictive
recursion algorithm \citep{newton1998nonparametric} which requires
only a single scan of all observations and avoids expensive MCMC. The
second approach is a two-step MCMC algorithm. It first distributes data
onto different machines and computes clusters locally in parallel. A second step
combines the local clusters into global clusters.  All steps are
carried out using MCMC simulation under a common DPM
model. \cite{ni2018scalable} extended the second approach to
non-conjugate models.

We propose a new CMC algorithm
specifically for aggregating subject-specific latent combinatorial
structures that often arise from BNP models. The proposed CMC
algorithm is applicable to both clustering and feature allocation
problems. It uses a similar notion, \textit{anchor points}, as in
\cite{kunkel2018anchored} but with a completely different purpose. In
\cite{kunkel2018anchored}, they used anchor points to address label switching in finite Gaussian mixture models whereas in this
paper, we use anchor points to combine posterior draws of
random clusters or features 
across different shards of the original data.
The proposed CMC can reduce the computation cost by at least
a factor of $S$ where $S$ is the number of computing cores at
disposal. Since modern high performance computing centers typically
have tens of thousands to hundreds of thousands computing cores in
total, $S$ is easily in the order of thousands.
%\vspace*{.15in}

The rest of this article is organized as follows. We provide relevant
background for CMC and two BNP models in Section \ref{sec:bg}. The
proposed CMC is introduced in Section \ref{sec:cmc}. The utility of
the proposed CMC is demonstrated with simulation studies in Section
\ref{sec:sim} and three real applications in Section \ref{sec:app}. We
conclude the paper by a brief discussion in Section \ref{sec:disc}.

\section{Background}
\label{sec:bg}

CMC schemes are algorithms to generate approximate posterior
Monte Carlo samples given a large dataset. The idea of CMC can be
summarized 
in three steps: (i) distribute a large dataset to multiple machines
(shards); (ii) run separate MCMC algorithm on
each machine without any inter-machine communication; and (iii)
aggregate Monte Carlo samples across machines to construct consensus
posterior samples. Step (iii) is non-trivial.

Let $\yb$ denote the full dataset and let $\yb_s$ denote shard $s$ for
$s=1,\dots,S$. And let $\thetab$ denote the global parameters. 
Assuming $\yb$ to be exchangeable, the posterior distribution can be
written as  
\begin{align}
\label{eqn:fp}
p(\thetab \mid \yb)\propto p(\yb \mid \thetab)p(\thetab)=\prod_{s=1}^Sp(\yb_s \mid \thetab)p(\thetab)^{1/S}
\end{align}
where the fractional prior preserves the total amount of prior
information. Expression
\eqref{eqn:fp} is exact and involves no approximation. Ideally, if one
can compute $p(\yb_s \mid \thetab)p(\thetab)^{1/S}$ analytically for each
shard $s$, then \eqref{eqn:fp} can be directly used to obtain
the posterior distribution based on the entire dataset. However,
$p(\yb_s \mid \thetab)p(\thetab)^{1/S}$ is usually not analytically
available and only Monte Carlo samples are returned from step (ii),
which necessitates step (iii).

Various methods have been proposed to aggregate Monte Carlo samples
such as weighted averaging \citep{scott2016bayes}, density
estimation \citep{neiswanger2013asymptotically,white2015piecewise},
geometric median \citep{minsker2014scalable} and a variational
algorithm \citep{rabinovich2015variational}. One common limitation
of existing approaches is that they cannot aggregate Monte Carlo
samples of subject-specific latent structure
(e.g. cluster assignment and feature allocation) that often arise in posterior inference under BNP models.
For example, in a feature allocation problem each shard
introduces new, additional feature memberships. That is, $\thb$ could
be partitioned as $\thb=(\thb_s, s=1, \ldots, S)$, with each
shard adding an additional set of parameters $\thb_s$.

% \subsection{Bayesian nonparametric models}
BNP models are  Bayesian models defined on an infinite-dimensional
parameter space. 
Common examples of BNP models include the DPM model for
clustering and the IBP prior for feature allocation. 
The  main attraction of BNP models are flexible modeling
and the full prior support in many models. 
For example, DP and IBP can
automatically select the number of clusters and features based on
available data. As examples illustrating the proposed CMC algorithm,
we consider applications of the DPM model and the IBP model in
three different inference problems. The inference problems are
clustering, using a DP prior; FA, using an IBP prior;
and double feature allocation (DFA, \citealt{Ni:19}), using an IBP prior.
For the implementation we use the
R packages \texttt{DPpackage} and \texttt{dfa}.
Next we introduce some notations by way of a brief review of the three
models and the inference problems.

\subsection{Dirichlet process mixtures and random partitions}
\label{sec:dpmrp}
Let $\yb_1,\dots,\yb_n$ denote data observed on experimental
units $i=1, \ldots, n$. Some applications call for clustering, i.e., a
partition of $[n] \equiv \{1, \ldots, n\}$ into
$[n] = \bigcup_{k=1}^K F_k$, with
$F_k \cap F_{k'}=\emptyset$ for $k \not= k'$.
A widely used model-based approach to implement inference on the
unknown partition is to assume i.i.d. sampling from a mixture model,
$\yb_i \sim \int f(\yb_i \mid \thb)\; dG(\thb)$. Here $f(\yb \mid
\thb)$ is, for example, a normal distribution with location $\thb$
(leaving the scale parameter as an additional hyperparameter), and
$G(\cdot) = \sum_h w_h \delta_{m_h}(\cdot)$ is a discrete mixing
measure.
Introducing latent variables $\thb_i$, the model can equivalently be
written as a hierarchical model
\begin{equation} % {gather}
\yb_i \mid \thb_i  \sim  f(\yb_i \mid \thb_i) \mbox{ and }
\thb_i  \sim  G.
% \nonumber \\  G   \sim  \DP(m, G_0).
\label{eq:mix}
\end{equation} % {gather}
The discrete nature of $G$ gives rise to ties among
the $\thb_i$, which in turn naturally define the desired partition of $[n]$.
Assume there are $K$ unique values, denoted by
$\thbs = \{\thbs_1, \ldots, \thbs_K\}$, and define clusters
$F_k = \{i:\; \thb_i=\thbs_k\}$. Sometimes it is more convenient to
alternatively represent the partition using cluster membership
indicators $s_i=k$ if $i\in F_k$. Yet another alternative representation, that will be useful
later, is using binary indicators $A_{ik}$ with $A_{ik}=1$ if $i \in
F_k$. Collecting all indicators in an $(n \times K)$ binary matrix
$\Ab=[A_{ik}]$, the constraint to non-overlapping subsets becomes
$\sum_k A_{ik}=1$ for all rows $i=1, \ldots, n$.  

Model \eqref{eq:mix} is completed with a BNP prior on $G$, for example,
$
G \sim \DP(m,G_0),
$
where $\DP(m, G_0)$ denotes a DP prior with concentration parameter $m$ and
base measure $G_0$. Model \eqref{eq:mix} with the DP hyperprior on $G$
defines the DPM model. See, for example, \cite{Ghoshal:10} for a good review.
The implied distribution $p(\Fb)$ over partitions $\Fb = \{F_1,
\ldots, F_K\}$ under a DPM model is known as the Chinese restaurant process.
Inference on the random partition is
straightforward to implement through, for example, Algorithm 8 in
\cite{neal2000markov}. Neal's algorithm 8 iterates between two steps:
% \begin{itemize}
% \item [i.] 
(i) For $i=1,\dots, n$, sample $s_i$ given the $i$th
observations $\yb_i$, cluster-specific parameters
$\thetab_1^\star,\dots,\thetab_K^\star$, and cluster assignments
$\bm{s}_{-i}$ for the rest of the observations.
% \item [ii.] 
(ii) For $k=1,\dots, K$, sample $\thetab_k^\star$ given data
$\yb_1,\dots, \yb_n$ and cluster assignments $s_1,\dots,s_n$.
%\end{itemize}

\subsection{Indian buffet process and feature allocation}
\label{sec:ibpfa}
The IBP is a BNP prior for random subsets 
$\Fb =\{F_k \subseteq [n];\; k=1, \ldots, K\}$ that can possibly overlap
and need not be exhaustive, i.e, without the restrictions of a
partition. Again, the number $K$ of random subsets is random.
The subsets are known as features. A prior
$p(\Fb)$ defines a random feature allocation model \citep{broderick2013cluster}.
Similar to before, we can alternatively represent the feature
allocation with feature membership indicators,
$A_{ik}=1$ if $i \in F_k$, now without the constraint to unit row sums.
The columns of $\Ab=[A_{ik}]$ represent the features $F_1, \ldots,
F_K$.

\cite{xu2015mad} use random feature allocation to develop inference
for tumor heterogeneity, i.e., the deconvolution of a heterogeneous
population of tumor cells into latent homogeneous subclones (i.e. cell subtypes).
The experiment records short reads counts of $n$ single nucleotide
variants (SNVs) (essentially, mutations relative to a given
reference)
in tumor tissues $j=1, \ldots, p$. 
The hypothesized homogeneous subclones are charecterized by the
presence or absence of these SNVs. In this application,
$A_{ik}=1$ if SNV $i$ is present in subclone $k$.
Let $y_{ij}$ denote the observed counts of SNV $i$
in sample $j$ and let $N_{ij}$ denote the total counts at locus $i$. Let $\ths_{jk}$ denote the unknown proportion of
subclone $k$ in tumor $j$. The experimental setup implies independent 
binomial sampling, for $i=1,\dots,n$ and $j=1,\dots,p$,
\begin{equation}
\label{eq:TH}
y_{ij} \sim \Bin(N_{ij}, p_{ij}) \mbox{~~with~~}
p_{ij} = b_jp_0+\sum_{k=1}^K \ths_{jk} A_{ik},
\end{equation}
where $p_0$ is the relative frequency of a SNV in the background and $\thbs_k=(\ths_{jk},\; j=1,\ldots,p)$ are
feature-specific parameters. The model is completed with a Dirichlet
prior on $(b_j, \ths_{j1},\ldots, \ths_{jK})$, a beta prior on $p_0$ and a prior on the feature
allocation $p(\Ab)$.

The most widely used prior $p(\Ab)$ for feature allocation is the IBP.
It defines a prior distribution for an
$(n\times K)$ binary matrix $\Ab=[A_{ik}]$  with a random number
of columns. 
We start the model construction assuming a fixed number $K$ of
features, to be relaxed later.  Conditional on $K$, $A_{ik}$'s are
assumed to be independent Bernoulli random variables, $A_{ik} \mid
\pi_k\sim Ber(\pi_k)$ with $\pi_k$ following a conjugate beta prior,
$\pi_k\sim Beta(m/K, 1)$. Here $m$ is a fixed
hyperparameter. Marginalizing out $\pi_k$,
$$
p(\Ab)= \prod_{k=1}^{K}
\frac{m\Gamma(c_k+\frac{m}{K})\Gamma(n-c_k+1)}
{K\, \Gamma(n+1+\frac{m}{K})},
$$
where $c_k=\sum_{i=1}^n A_{ik}$ is the sum of the $k$th column of
$\Ab$.

Let $H_n=\sum_{i=1}^n1/i$ be the $n$-th harmonic number.  Next, take
the limit $K\rightarrow \infty$ and remove columns of $\Ab$ with all
zeros. Let $K^+$ denote the number of non-empty columns.  The
resulting matrix $\Ab$ follows an $\IBP(m)$ prior (without a
specific column ordering), with probability
\begin{equation}
\label{ibp} p(\Ab)= \frac{m^{K^+}\exp\{-m H_n\}} {K^+!}
\prod_{k=1}^{K^+}\frac{\Gamma(c_k)\Gamma(n-c_k+1)} {\Gamma(n+1)}.
\end{equation} With a finite sample size, the number $K^+$ of
non-empty columns is finite almost surely.  Let $r_{-i,k}$ denote the
sum in column $k$, excluding row $i$.  Then the conditional
probability for $A_{ik}=1$ is
\begin{align}
  \label{eqn:pc} p(A_{ik}=1 \mid \Ab_{-i,k})=r_{-i,k}/n,
\end{align} provided $r_{-i,k}>0$, where $\Ab_{-i,k}$ is the $k$-th
column of $\Ab$ excluding $i$-th row.  And the distribution of the
number of new features (non-empty columns) for each row is $\Poi(m/n)$.
Posterior inference can be carried out using an algorithm similarly to Neal's
algorithm 8. The posterior distribution may present many peaked modes for moderate to large total counts $N_{ij}$, which makes MCMC inefficient.  To improve mixing, \cite{ni2019parallel} used parallel tempering to flatten the posterior while still targeting the right posterior distribution. We will follow their strategy in MCMC.
\subsection{Double feature allocation}
\label{sec:dfadfa}

\begin{figure}
	\begin{center}
		\includegraphics[width=.3\textwidth]{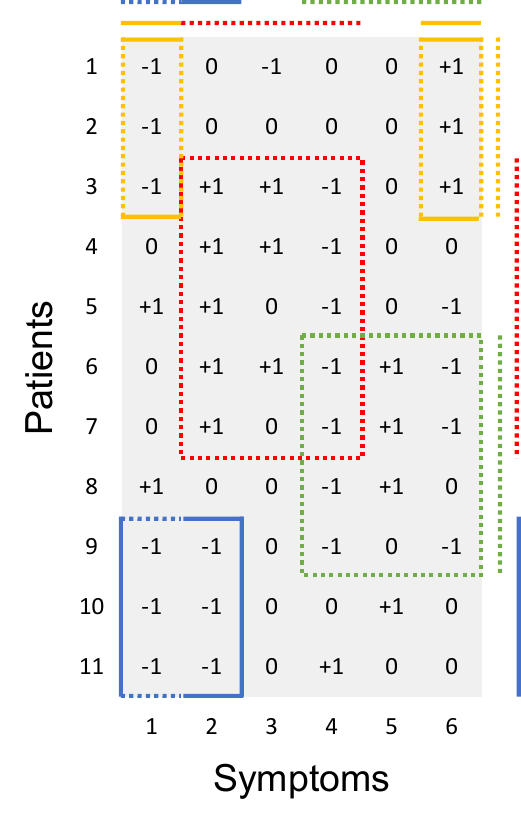}
	\end{center}
	\caption{Double feature allocation: simultaneous feature allocation
		on rows (patients) and columns (symptoms) defines diseases.
		The data are recorded symptoms, $j=1, \ldots, p$, for patients,
		$i=1, \ldots, n$, in electronic health records (EHR) data.
		Each subset $F_k$ of patients and matching subset $R_k$ of symptoms is a
		different disease, marked by different boxes. 
		The data $(n \times p)$ matrix $\Yb=[y_{ij}]$
		records trinary symptoms with possible outcomes $\{-1, 0, 1\}$.}
	\label{fig:dfa}
\end{figure}

Some applications call for simultaneous clustering of rows and columns
of a data matrix. This is known as bi-clustering \citep{hartigan1972direct}. \cite{Ni:19} introduces a similar prior model for random row and
column subsets, but now without the restrictions of a partition, i.e.,
random feature allocation on rows and columns simultaneously.
Figure \ref{fig:dfa} illustrates how matching pairs $(F_k,R_k)$
of subsets $F_k \subseteq [n]$
and $R_k \subseteq [p]$ define a disease in electronic health records
(EHR) data. The data is an $(n \times p)$ matrix $\Yb=[y_{ij}]$ of
recorded symptoms, $j=1, \ldots, p$, on patients $i=1, \ldots, n$.
A prior on random pairs $(F_k,R_k)$ of matching subsets
defines a double feature allocation (DFA, \citealt{Ni:19}).
A simple extension of a feature allocation model, such as the IBP, can
be used to define a DFA.
Let $\Ab$ denote a matrix of feature membership indicators for the
subsets $F_k \subseteq [n]$, as before. We first assume
$\Ab \sim \IBP(m)$.
In particular, $p(\Ab)$ induces a prior on the number of
subsets $K$. Conditional on $\Ab$ we then define a second
membership matrix $\Cb$, for membership in matched subsets $R_k
\subseteq [p]$.
Here $\Cb=[C_{jk}]$ with $C_{jk}$ recording membership of column $j$
in $R_k$. In anticipation of the upcoming sampling model
for trinary symptoms, $y_{ij} \in \{-1,0,1\}$, 
we allow for membership $C_{jk} \in \{-1, 0, 1\}$, with $-1$ indicating that
disease $k$ favors symptom $j$ at level $-1$ (e.g., low blood pressure),
$1$ indicating that the symptom is favored at level $1$ (e.g., high
blood pressure), and $0$ indicating that the symptom is not related to
disease $k$.
Let $\pib=(\pi_{-1},\pi_0,\pi_1)$ denote a probability vector.
We assume $p(C_{jk} = c) = \pi_c$, with a
conjugate hyperprior $\pib \sim Dir(a_{-1},a_0,a_1)$.
% Let $\Cb_j$ denote the $j$-th row of $\Cb$, and let
% Let $C_{jk}^+ = I(C_{jk}=+1)$ and $C_{jk}^-=I(C_{jk}=-1)$.
We assume conditionally independent trinary latent logistic regression as a
sampling model for the observed symptoms 
$y_{ij} \in \{-1, 0, 1\}$ for patient $i$,
\begin{equation}
\label{eq:cat}
p(y_{ij}=y \mid \Ab, \thbs, \etab) \propto
\begin{cases}
e^{\eta_j^-+ \sum_{k=1}^K w_{jk}^-\, I(A_{ik}=1, C_{jk}= -1)}
&  \mbox{if~~}y= -1\\
1 & \mbox{if~~} y= 0\\
e^{
	\eta_j^++ \sum_{k=1}^K w_{jk}^+\, I(A_{ik}=1, C_{jk}=1)}
&  \mbox{if~~}y=1
\end{cases}
\end{equation}
where 
$(w_{jk}^+, w_{jk}^-,\; j=1,\ldots,p)$ 
are feature-specific weights, and
$\etab=(\eta_j^-,\eta_j^+)$ are symptom-specific offsets.
The model is completed with priors for the hyperparameters, $\eta_j^-$,
$\eta_j^+\sim N(0,\tau^2)$ and $w_{jk}^-$, $w_{jk}^+\sim
Ga(1,\tau_w)$. 
In model \eqref{eq:cat} the feature-specific parameters are
$\thbs_k = (C_{jk}, w_{jk}^+, w_{jk}^-;\; j=1, \ldots, p)$ for feature
$F_k$. 
Recognizing the columns of $\Cb$ as just another part of
feature-specific parameters $\thbs_k$ reveals the nature of the DFA
model as a special case of a feature allocation, with one of the
feature-specific parameters selecting a matching subset of the columns
in the data matrix. 

Straightforward modifications define similar models for
categorical data with fewer or more categories.  Posterior
inference can be carried out using an algorithm similarly to Neal's
algorithm 8; see \cite{Ni:19} for implementation details.

\section{A consensus Monte Carlo algorithm 
	for random subsets}
\label{sec:cmc}
% \subsection{Algorithm}
We describe the proposed CMC algorithm in its
general form. Let $y_1,\dots,y_n$ denote the data of $n$ observations. 
Let $\Ab=[A_{ik}]$ denote the  latent subset membership
matrix (for clusters, features, or row-features in random
partitions, FA, and DFA, respectively) where $A_{ik}=1$ if observation $i$ belongs to subset $k$. Let $\xib_i$ be the $i$-th row of $\Ab$.
Let $\thbs=\{\thbs_1,\thbs_2,\cdots\}$
denote an infinite sequence of subset-specific parameters and let
$\thb_{\Ab_i}$ denote a subsequence indexed by $\Ab_i$,
i.e. $\thb_{\Ab_i}=\{\thbs_k \mid A_{ik}=1\}$. 
Many BNP models
including DPM, FA, and DFA models can be
generically written as a hierarchical model,
\begin{eqnarray}
y_i \mid \thbs, \xib_i & \ind &  p(y_i \mid \thb_{\Ab_i}) \nonumber \\
\thbs_k \mid \Ab       & \iid & \pi_\th(\cdot)          \nonumber \\
\xib= [\xib_1\;  \cdots \xib_n]' & \sim & \pi_{\Ab}(\cdot)
\label{eq:bnp}
\end{eqnarray}
for $i=1,\dots,n$ and $k=1,2, \ldots$.
The sampling model for $y_i$ in the first line depends on the specific
inference model. In the case of a random partition it is the kernel of
the mixture model in \eqref{eq:mix}.
In the case of an FA or DFA model the sampling model
might include multiple $\thbs_k$ when $\sum_k A_{ik}>1$, as in
\eqref{eq:TH} or \eqref{eq:cat}. The specific interpretation is
problem-specific. 
The dependence on $\Ab$ in the second line is only indirect
through the random number of columns in $\Ab$ which defines the number
of subsets. 
In DPM models,
$\pi_\theta(\cdot)=G_0$ is the baseline distribution and
$\pixi(\cdot)$ is the Chinese restaurant process. In FA and DFA models,
$\pi_\theta(\cdot)$ is the prior of feature-specific parameters and hyperparameters, and
$\pixi(\cdot)$ is the IBP. 
% Recall that $\xib$ can be equivalently
% written as $\Pib=\{F_1,\dots, F_K \}$ for which $i\in F_k$ if and
% only if $A_{ik}=1$.
% distributions on $C_{jk}$, $w_{jk}^-$, $w_{jk}^+$, $\eta_j^-$ and $\eta_j^+$

For a small to moderate sample size $n$, MCMC has been commonly used
to implement posterior inference. However, when $n$ is large, MCMC
becomes computationally prohibitive because at each iteration, it has to scan
through all observations. The idea of CMC is to distribute the large
dataset onto many shards so that MCMC can be efficiently implemented
on each shard with much smaller sample size. Let $S$ be a large
integer. We randomly divide the 
observations into $(S+1)$ non-overlapping shards $\ib_s\subset
\{1,\dots,n\}=\cup_{s=1}^{S+1}\ib_s$ and $\yb_s=\{y_i: \; i\in \ib_s
\}$. 
Define $\ybt_s=\yb_s\cup\yb_{S+1}$ for $s=1,\dots,S$, so that $\{\ybt_s \}$ are new shards that all share $\yb_{S+1}$ but are otherwise
disjoint. We call $\yb_{S+1}$ the anchor points. Let
$\{\thb_s^{(t)},\Pib_s^{(t)}\}_{t=1}^T=MCMC(\ybt_s)$
where $\Pib_s^{(t)}=\{F_{sk}^{(t)} \}_{k=1}^{K_s^{(t)}}$ denote $T$
Monte Carlo samples obtained from the MCMC algorithm applied  to the
shard $\ybt_{s}$. To aggregate the Monte Carlo samples
from shards $s$ and $s'$, we consider merging (\bch i.e. taking the union
of\ech) two clusters or features $F_{sk}^{(t)}$ and $F_{s'k'}^{(t)}$
if  
\begin{equation}
d_{sk,s'k'}^{(t)}=
\frac{D(F_{sk}^{(t)},F_{s'k'}^{(t)})}
{C(F_{sk}^{(t)},F_{s'k'}^{(t)})+D(F_{sk}^{(t)},F_{s'k'}^{(t)})}
<\eps,
\label{eq:dFkk'}       
\end{equation}
where $D(F_{sk}^{(t)},F_{s'k'}^{(t)})$ is the number of different
elements in $F_{sk}^{(t)}\cap \ib_{S+1}$ and $F_{s'k'}^{(t)}\cap
\ib_{S+1}$ and $C(F_{sk}^{(t)},F_{s'k'}^{(t)})$ is the number of
common elements. By convention, we set $d_{sk,s'k'}^{(t)}=1$ if $C(F_{sk}^{(t)},F_{s'k'}^{(t)})=D(F_{sk}^{(t)},F_{s'k'}^{(t)})=0$. In words, if two random subsets (clusters or
features) have similar
sets of anchor points, we merge them. The similarity is controlled by
the tuning parameter $\eps$ which is a small fixed constant.
\bch Consider a toy example with 8 data points $\{1,\dots,8\}$. We
assign $\{1,2\}$ to shard $s=1$, $\{3,4\}$ to shard $s=2$, and
$\{5,6,7,8\}$ as anchor points. Suppose $F_{11}=\{1,5,6,7,8\}$,
$F_{12}=\{2,8\}$, $F_{21}=\{3,4,5,6,7 \}$ and
$F_{22}=\{3,6,8\}$. Using $\epsilon=0.3$, we will merge $F_{11}$ and
$F_{21}$  into $F^\star=F_{11}\cup F_{21}=\{1,3,4,5,6,7,8\}$, but
keep $F_{12}$ and $F_{22}$ unchanged. \ech 
The choice of \eqref{eq:dFkk'} includes arbitrary
choices. In particular, we note that, for example, the IBP includes
positive prior probability for two identical columns in $\Ab$, which
could question the appearance of identical subsets of anchor points as
a criterion for merging. However, in most applications, including the
two motivating applications related to feature allocation in this
article, this is not a desirable feature of the IBP, and we argue
that the criterion introduces an even desirable approximation.
Alternatively, the criterion could include a comparison of
feature-specific parameters $\thbs_k$. In simulation studies and the
motivating applications we found the proposed criterion to work well,
and prefer the simplicity of \eqref{eq:dFkk'}. We also find that the proposed algorithm is relatively robust with respect to the choice of
$\eps$ (see Section \ref{sec:mnist} for sensitivity analysis) if the number $|\ib_{S+1}|$ of anchor points is
sufficiently large.

If we decide to merge $F_{sk}^{(t)}$ and $F_{s'k'}^{(t)}$, we merge
the associated parameters $\thb_{sk}^{\star(t)}$ and
$\thb_{s'k'}^{\star(t)}$ according to the following operation.  For categorical
parameters, such as column $k$ of $\Cb_{jk}$ in \eqref{eq:cat}, a majority
vote is used, weighted by
the relative sizes of the two merged random subsets. 
For continuous parameters that index a sampling model, such
as in \eqref{eq:mix}, the values are averaged, again based on the
relative subet sizes. \bch The weighted average is
valid  only when the support of the parameter is a convex set, which
is true in most applications including all our examples.  
When the support is not convex (e.g. weighted directed acyclic
graphs), alternative merging operations need to be designed on
a case-by-case basis. \ech 
% by taking the average for continuous parameters
% weighted by the sizes of $F_{sk}^{(t)}$ and $F_{s'k'}^{(t)}$, and
% taking the majority vote for categorical parameters.
Note that the aggregating step is trivially parallelizable with
respect to the number $T$ of Monte Carlo samples and hence the computation time is negligible compared to MCMC. The complete CMC is
summarized in Algorithm \ref{alg:1}.
\bch In the algorithm, {\em parfor} indicates a parallel loop, while
{\em for} indicates a sequential loop. \ech
\begin{algorithm}
	\caption{}
	\label{alg:1}
	\begin{algorithmic}[1]
		% \Function{SIGN}{$\bm{y},R$}
		\State \textbf{Data preparation.} Split data into $S+1$ disjoint
		shards   $\ib_s\subset \{1,\dots,n\}=\cup_{s=1}^{S+1}\ib_s$. Form
		$S$ shards: $\ybt_s=\yb_s\cup\yb_{S+1} $ for $s=1,\dots,
		S$. Anchor points $\yb_{S+1}$ are present in every shard, whereas
		$\yb_s$ only appear in shard $s$.  
		\ParFor{$s=1,\dots,S$ }
		\State $\{\thb_s^{(t)},\Pib_s^{(t)} \}_{t=1}^T=MCMC(\ybt_s)$
		\EndParFor
		\ParFor{$t=1,\dots,T$}
		%\For{each pair of pairs $(s,k)$ and $(s',k')$}
		\State \bch randomly  permutate the order of the shards\ech
		\For{\bch each pair of pairs $(s,k)$ and $(s',k')$, $s'>s$\ech}
		\If{$d_{sk,s'k'}^{(t)}<\eps$}
		\State merge $F_{sk}^{(t)}$ and $F_{s'k'}^{(t)}$
		\State merge $\thb_{sk}^{\star(t)}$ and $\thb_{s'k'}^{\star(t)}$
		%\State \bch break the loop over $k'$ (optional)\ech
		\EndIf
		\EndFor
		\EndParFor
		\State \textbf{Output:} $\{\thb^{(t)},\Pib^{(t)} \}_{t=1}^T$
		
	\end{algorithmic}
\end{algorithm}
We implement the algorithm in the upcoming simulation studies, and in the
motivating applications in the context of DPM models and FA and DFA based on IBP models.
However, the approach is more general. It remains valid for any
alternative BNP prior on $G$ in \eqref{eq:mix}, and any alternative
FA prior.
For example, the BNP prior could be any other random discrete
probability measure, including a normalized completely random measure
as in
\cite{barrios&al:13,favaro&teh:13} or \cite{Argiento:2010}. The IBP could be replaced by any other random feature allocation
\citep{broderick2013feature}. Also, the sampling models in \eqref{eq:mix}, \eqref{eq:TH} and
\eqref{eq:cat} are examples. Any other sampling model could be
substituted, including a regression on additional covariates, or, in
the case of feature allocation, a linear-Gaussian model. While we use it here only for BNP models, the same algorithm
can be implemented for inference under any parametric model for random
subsets, for example, finite mixture models or finite feature allocation models. 
%See \cite{Ni:19} for a brief discussion of alternative models. \ech

We propose a simple diagnostic to summarize the level of approximation
in the CMC.  First select any two shards, without loss of generality
assuming they are $\ib_1$ and $\ib_2$.  We then apply CMC with
$\ib_{S+1}$ as anchor points. Denote the point estimate of random
subsets (e.g. clusters, latent features) by
$\widehat{\Ab}_{\mbox{\tiny CMC}}$.  In addition, we run a full MCMC
simulation in $\ib_1\cup \ib_2\cup \ib_{S+1}$ and denote the point
estimate by $\widehat{\Ab}_{\mbox{\tiny MCMC}}$. We summarize the
level of the approximation by measuring the distance between
$\widehat{\Ab}_{\mbox{\tiny MCMC}}$ and $\widehat{\Ab}_{\mbox{\tiny
		CMC}}$. We illustrate the diagnostic in Section \ref{sec:mnist}.

\section{Simulation}
\label{sec:sim}
We carry out simulation studies to assess the proposed CMC algorithm
for DPM, FA and DFA models. We use relatively small
datasets in the simulations, so that we can make comparison with full
MCMC. Scalability will be explored later, in applications.  For all
models, we evenly split the observations into 5 shards and use one of
the shards as anchor points. We report frequentist summaries based on
50 repetitions.  For both CMC and MCMC, we run 5,000 iterations,
discard the first 50\% of Monte Carlo samples as burn-in and only keep
every 5th sample. We choose $\epsilon=0.1$ and find it work well throughout the simulations and applications. 
The sensitivity of the choice of $\eps$ will be assessed in Section \ref{sec:mnist}.

\subsection{Simulation 1: Clustering under the DPM model}
The first simulation considers a CMC approximation of posterior
inference in a DPM model for a $p=4$ dimensional variable $\yb_i$, \bch $i=1,\dots,n$, and sample size $n=1,000$\ech: 
% To support the earlier claim in Section \ref{sec:ppmx} that SIGN can
% be applied with a range of BNP mixture models, we first carry out
% simulations to demonstrate the effectiveness of SIGN applied to
% \PYM. The probability model of \PYM is given by 
\begin{eqnarray*}
	\yb_i \mid \mub_i,\Sigmab_i{\buildrel\rm ind\over\sim} p(\yb_i \mid \mub_i,\Sigmab_i),\mbox{~~~~}\mub_i,\Sigmab_i \mid G{\buildrel\rm iid\over\sim} G,\mbox{~~~~} G\sim DP(m,G_0), 
\end{eqnarray*}
where
$G_0(\mub,\Sigmab)=
N_p(\mub \mid \bm{0},\Sigmab/\kappa_0)\times IW(\Sigmab \mid
b,\bm{I}_p)$. The hyperparameters are 
$m=1,\kappa_0=0.01$ and $b=p$.

% We consider $p=5$ variables and
We construct a simulation truth with $K=4$ true clusters with equal
sizes.  For $i=1, \ldots, n$, we generate data
$\yb_i \mid s_i=k \sim N_p(\mub_k,\Sigmab_k)$, 
where $\mub_1=(-1,1,-1,1)^T$, $\mub_2=(1,-1,1,-1)^T$,
$\mub_3=(-1,-1,1,1)^T$, 
$\mub_4=(1,1,-1,-1)^T$, and
$\Sigmab_k=0.4I_4$, for $k=1,\dots,4$.
%$\Sigmab_2=\diag(1.25^2,0.1,1,1,1)$,
%$\Sigmab_3=\diag(1,1,1,0.1,0.25)$, 
%\[\Sigmab_4=\blkdiag\left(\left[\begin{array}{cc}0.1&0.05\\0.05&0.1\end{array}
%  \right],I_3\right)\mbox{~and~}
%\Sigmab_5=\blkdiag\left(I_3,\left[\begin{array}{cc}0.25&0.125\\0.125&0.25\end{array}
%  \right]\right).\]
% Here $\diag(\cdot)$ and $\blkdiag(\cdot)$ denotes a diagonal
%matrix and a block diagonal matrix, respectively, with the given
%diagonal elements. 
%In words, clusters 1, 2, and 4 are characterized by a shift in
%the distribution for the first two  variables  $y_{i1}$
%and $y_{i2}$ with different correlation 
%structures, whereas clusters 3 and 5 are characterized by a shift
%in the third and fourth  variables  $y_{i4}$ and
%$y_{i5}$. And $y_{i3}$ plays the role of a ``noisy''  variable
%with a common distribution across all clusters.
%Variables  that do not  characterize  clusters (such as
%$y_{i3},y_{i4},y_{i5}$ in clusters 1, 2 and 4) are independently
%sampled from standard normal distributions. 
\bch The scatter plot of one randomly selected simulated data
  set is shown in  Figure S1(a) in the Supplementary
Materials. \ech

In Figures \ref{fig:dpmcmc} and
\ref{fig:dpmmcmc}, we show the bar plots of posterior modes
$\widehat{K}$ for the number of clusters across repeat simulations, evaluated using the CMC and a full MCMC implementation,
respectively. 
%CMC correctly identifies the 
%number of clusters in 37 out of 50 simulations, \bch matching the full
%MCMC. \ech
Compared to full MCMC, it tends to slightly overestimate the number of
clusters. This might be due to the fact that inference under the DP
prior typically includes many small clusters, \bch which  are
likely to include few or no anchor points and are therefore unlikely
to be merged with other clusters. 
 This is less of a problem under prior partition models other
  than the DPM, for example, a finite mixture model that encourages
  more balanced cluster sizes. \ech
% mitigate this issue, one
% can replace the DPM by a finite mixture model which encourages more
% balanced cluster sizes.  \ech
%  Also, posterior simulations in each
% shard are carried out independently. Due to posterior uncertainty in
% including anchors in parts of what could be the same cluster across
% two shards, the algorithm might not recognize this. 
We report the
misclustering rates $e_{\Ab}$ and the MSEs of parameter estimation
$e_\theta$ in Table \ref{tab:ocdfa2}. 
% The performance of CMC stays virtually the same for $\eps \in [0.1,0.5]$. 

\begin{figure}
	\centering
	\subfigure[DPM-CMC]{\includegraphics[width=.34\textwidth]{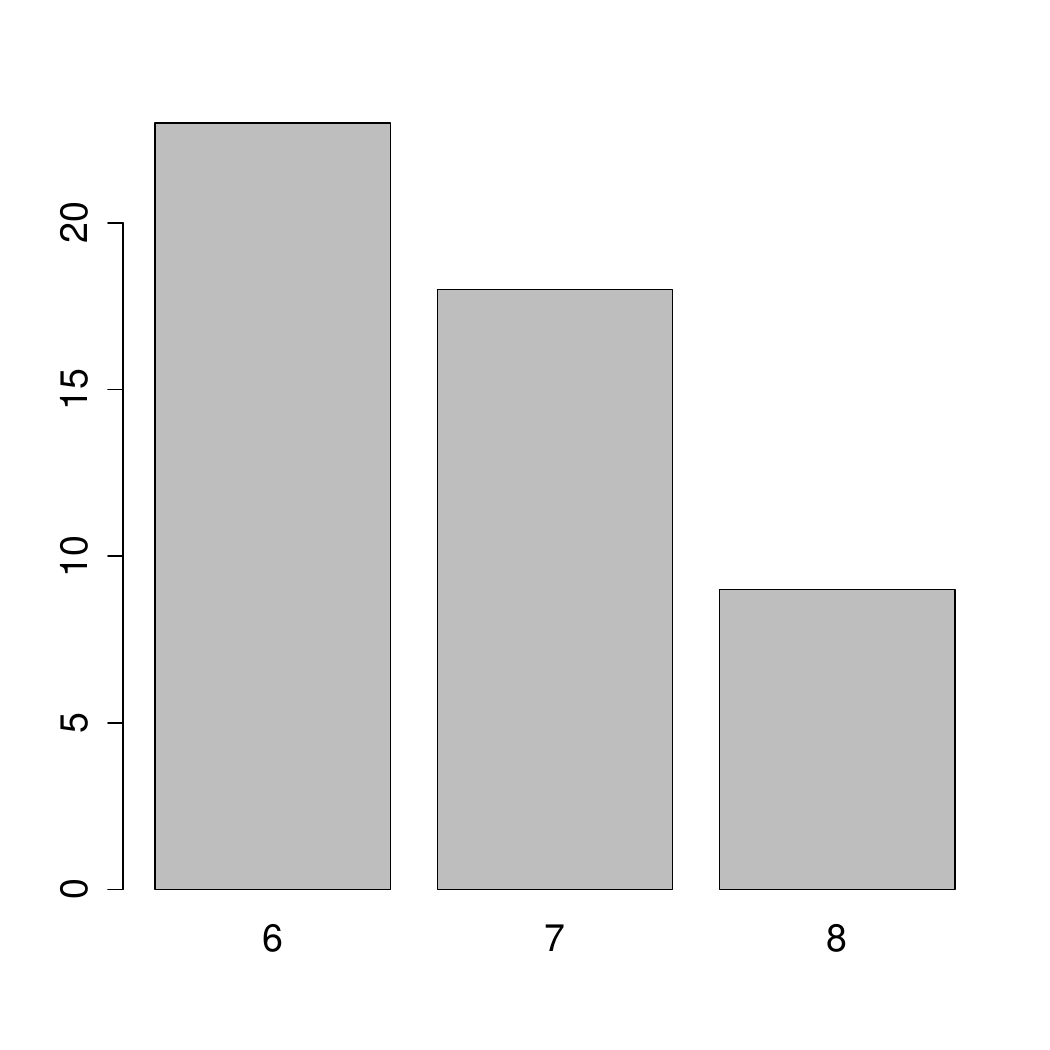}\label{fig:dpmcmc}}	\subfigure[DPM-MCMC]{\includegraphics[width=.34\textwidth]{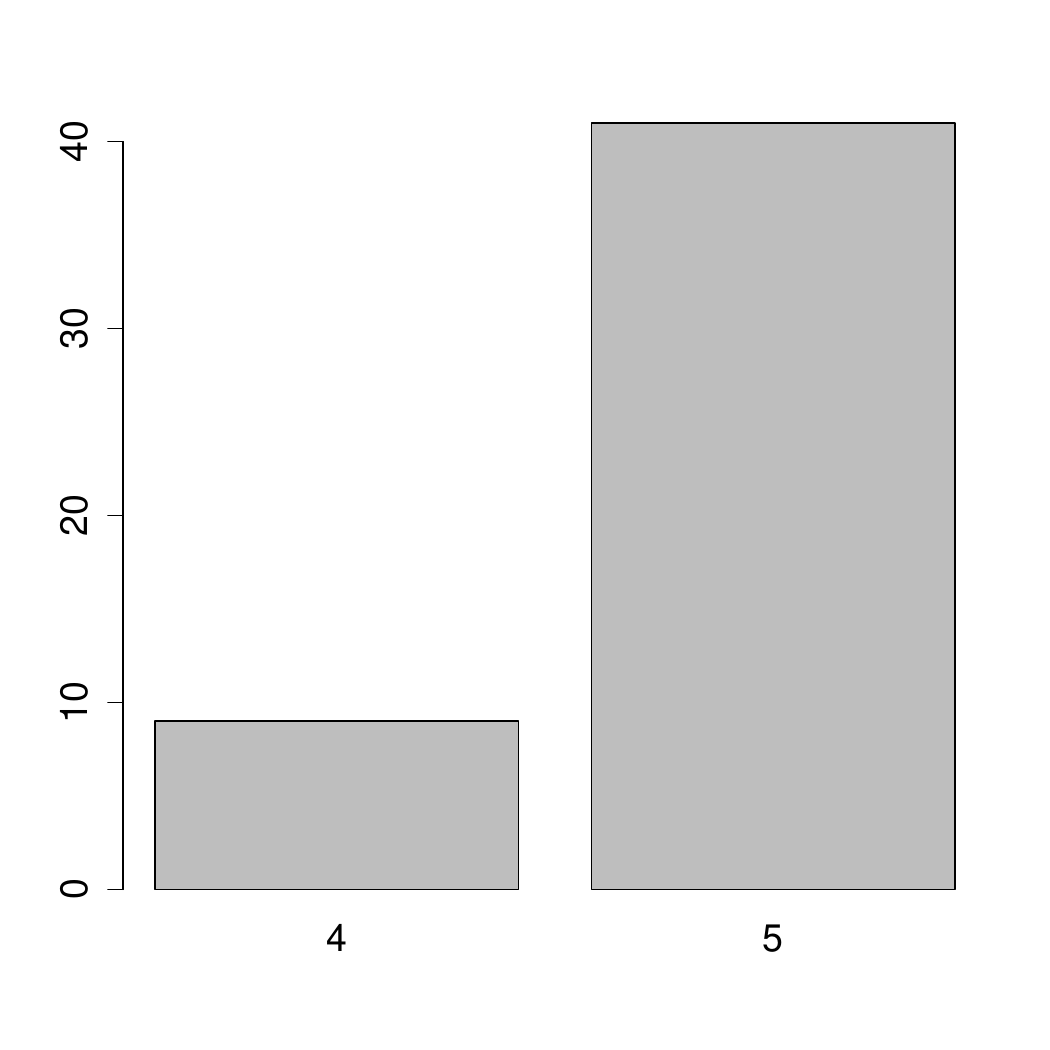}\label{fig:dpmmcmc}}	\subfigure[FA-CMC]{\includegraphics[width=.34\textwidth]{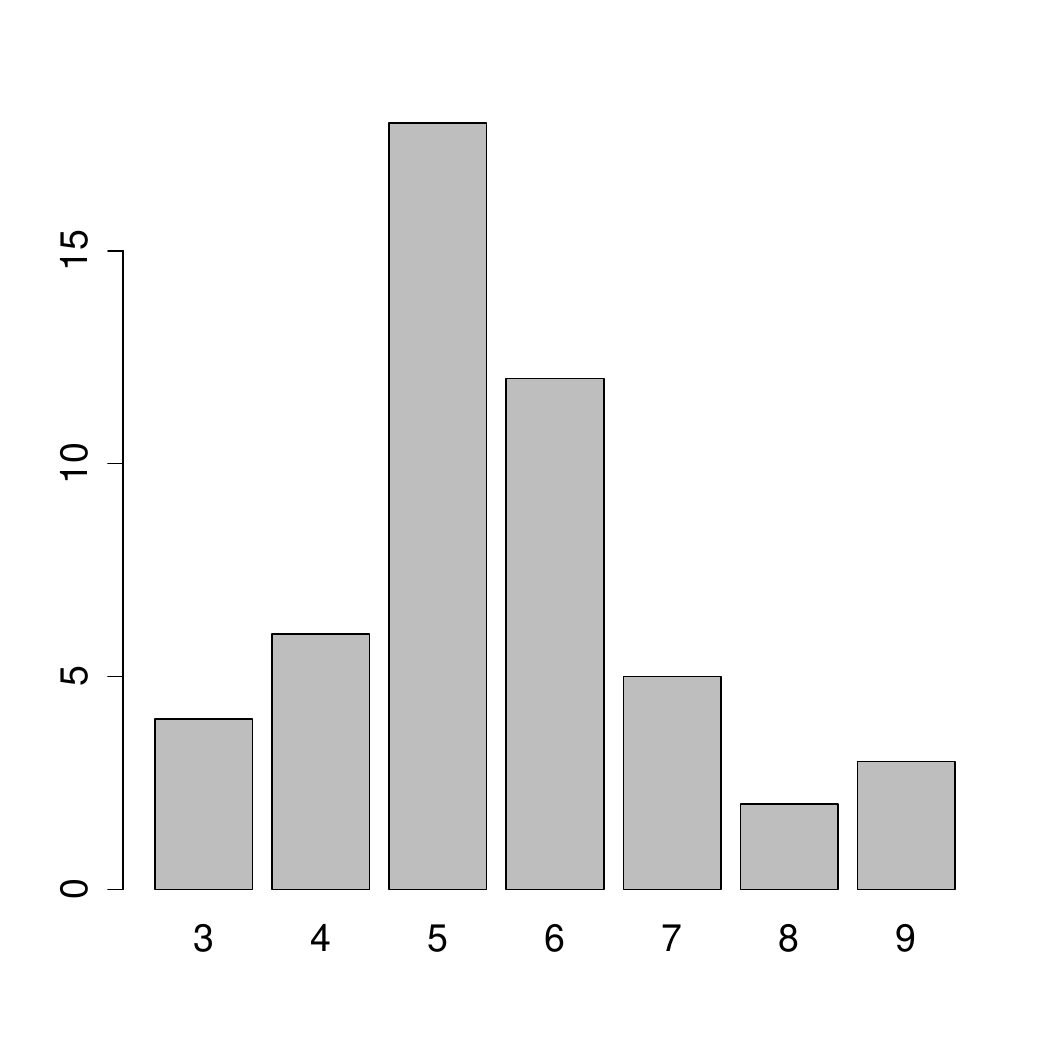}\label{fig:facmc}}
	\subfigure[FA-MCMC]{\includegraphics[width=.34\textwidth]{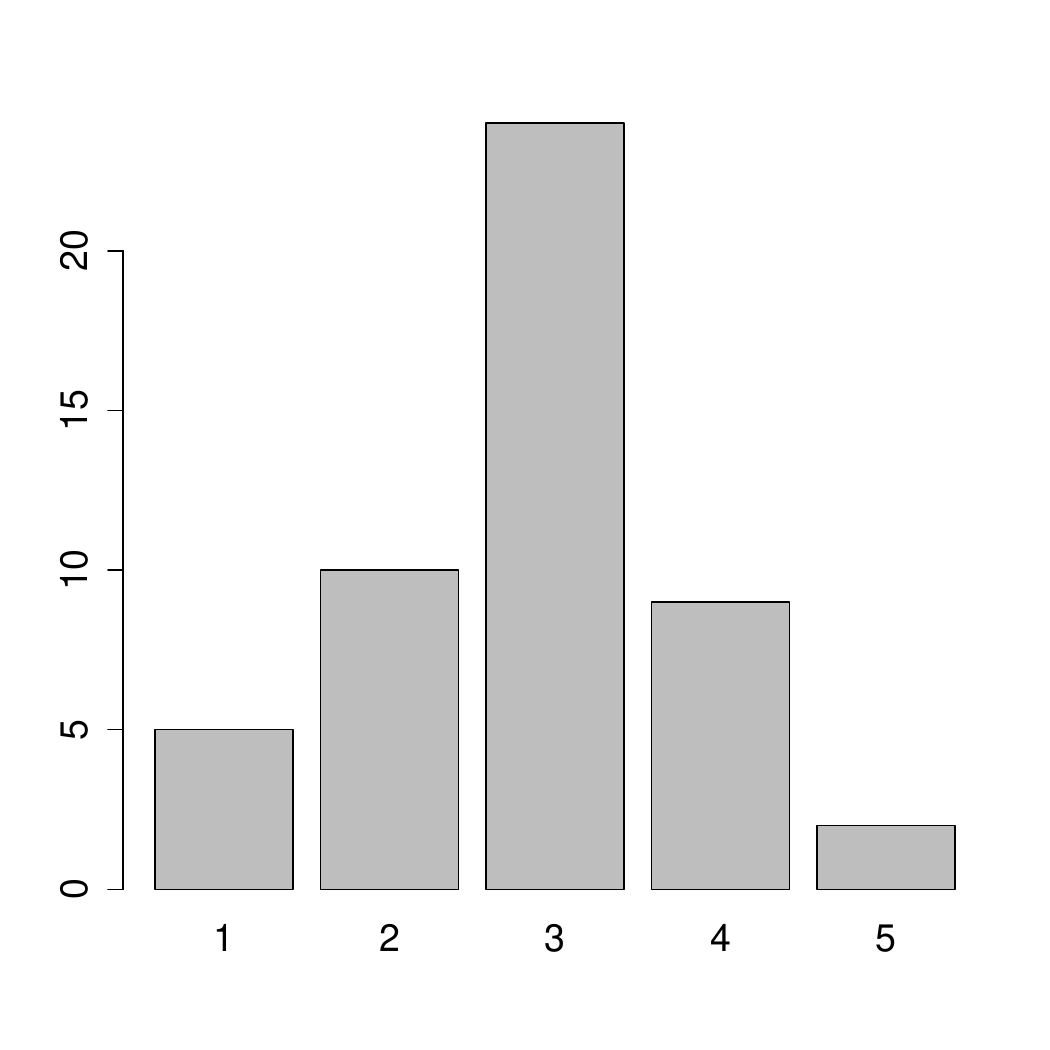}\label{fig:famcmc}}
	\subfigure[DFA-CMC]{\includegraphics[width=.34\textwidth]{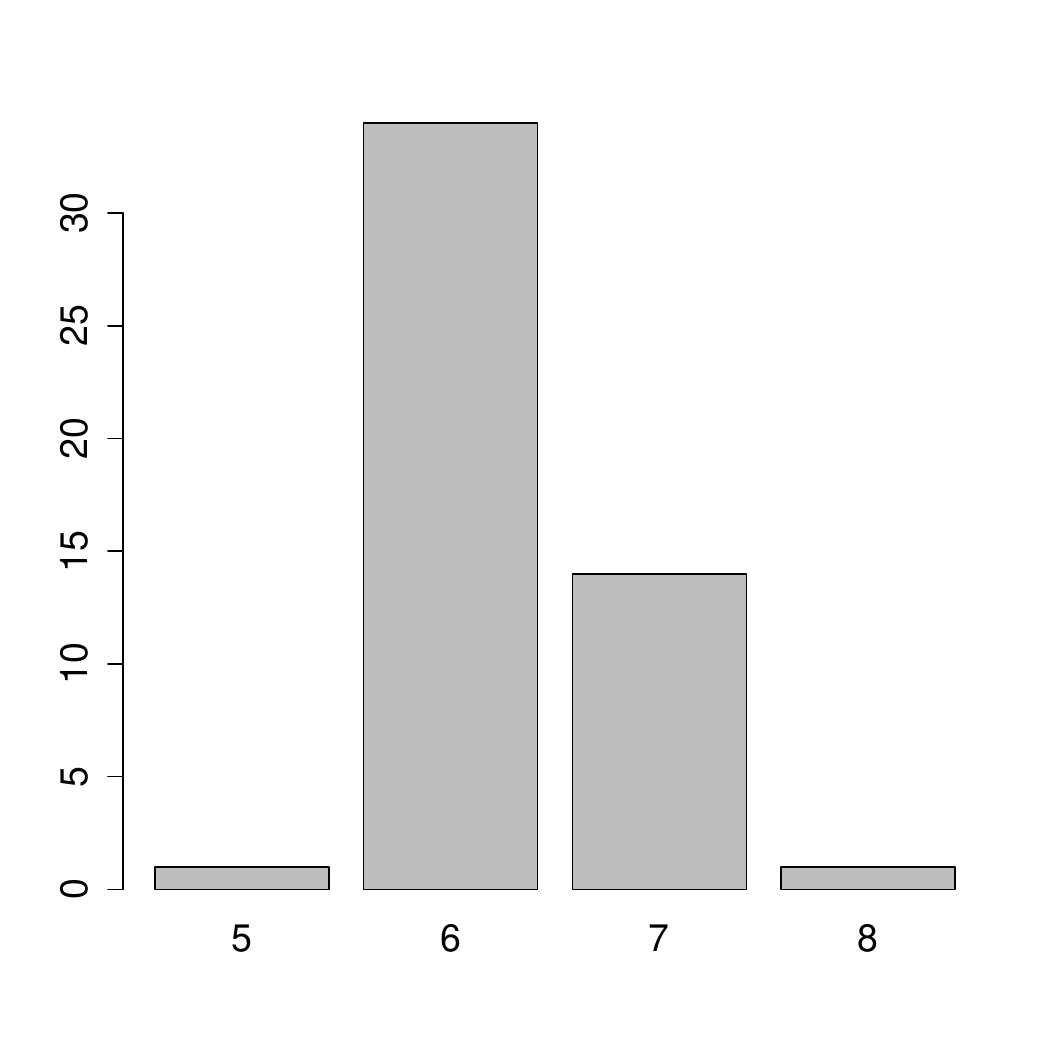}\label{fig:dfacmc}}
	\subfigure[DFA-MCMC]{\includegraphics[width=.34\textwidth]{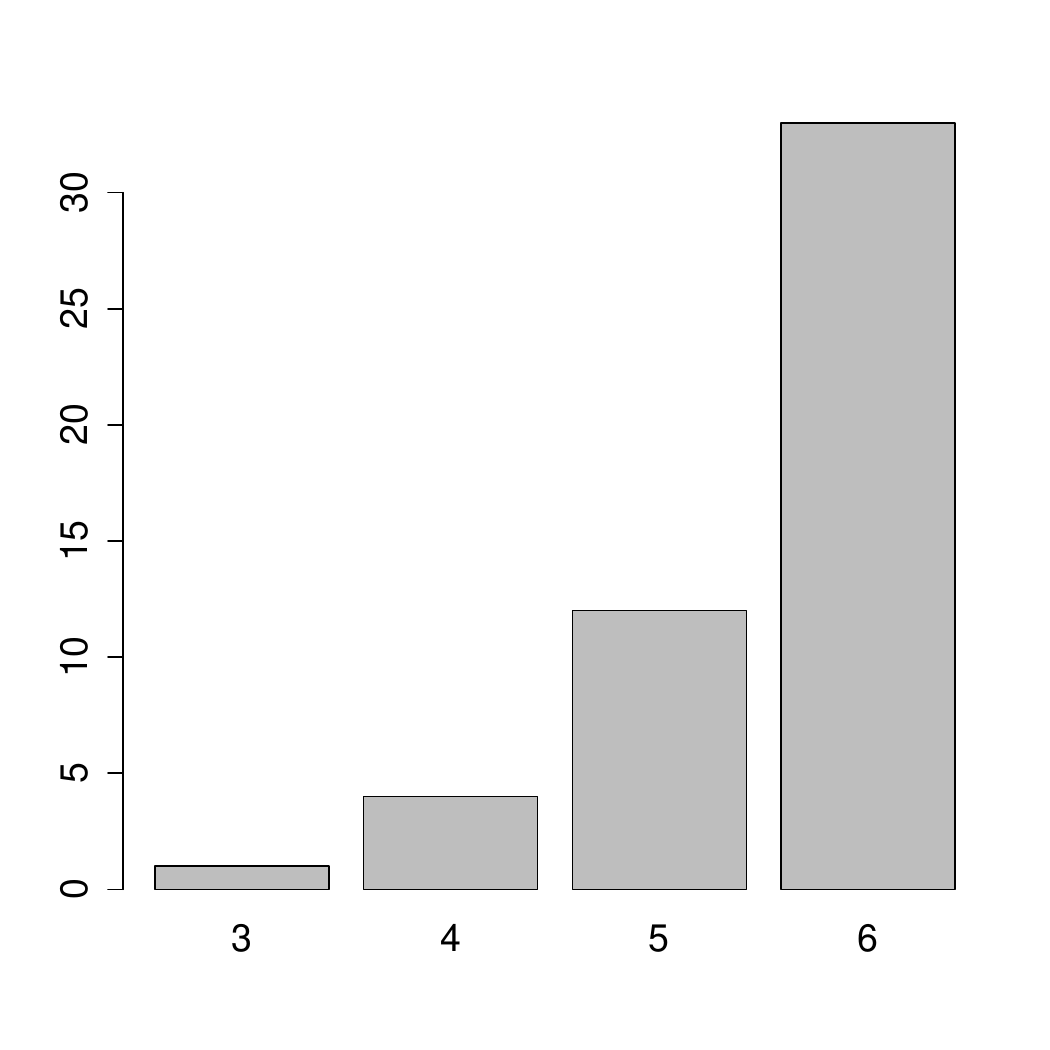}\label{fig:dfamcmc}}
	\caption{Simulations 1, 2 and 3:
		Bar plots of posterior mode $\widehat{K}$ across
		repeat simulations for CMC and full MCMC implementations for Simulations 1 (a and b), 2 (c and d), and 3 (e and f), respectively.} 
	\label{figure:spr}
\end{figure}

%\begin{table}	
%\caption{\bch Simulation 1, 2 and 3.  \ech
%  We report the performance of CMC versus full
%  MCMC for three models, DPM, FA and DFA.  The error  $e_{\Ab}$
%  reports the misallocation rate in estimating $\Ab$, and $e_\theta$ reports the MSE or average Hamming distance in estimating 
%  subset-specific  continuous parameters (simulation 1 and 2) or
%  categorical matrix parameters (simulation 3). The standard 
%  deviations are given within the parentheses. 
%   Both, mean and standard deviation, are with respect to repeat
%  simulations. 
%  Varying $\eps$ from 0.1 to 0.5 gives virtually the same performance.}  
%	\centering
%	\begin{tabular}{cccccccccc}
%		\\\hline\hline
%		&\multicolumn{2}{c}{DPM}&
%		&\multicolumn{2}{c}{FA} &
%                &\multicolumn{2}{c}{DFA}\\
%		\cline{2-3} \cline{5-6} \cline{8-9}
%		&CMC&MCMC&&CMC&MCMC&&CMC&MCMC \\
%		$e_{\Ab}$&
%0.13 (0.07)&0.05 (0.04)&& 0.14 (0.04) & 0.05 (0.06) && 0.06 (0.01)&0.02 (0.00)\\
%		$e_\theta$&
%0.02 (0.03)&0.01 (0.01)&& 0.01 (0.01) & 0.02 (0.03) &&0.06 (0.07)&0.02 (0.01)\\
%		\hline
%	\end{tabular}
%	\label{tab:ocdfa2}
%\end{table}

\begin{table}	
	\caption{Simulations 1, 2 and 3.
		We report the performance of CMC versus full
		MCMC for three models, DPM, FA and DFA.  The error  $e_{\Ab}$
		reports the misallocation rate in estimating $\Ab$, and $e_\theta$ reports the MSE or average Hamming distance in estimating 
		subset-specific  continuous parameters (simulations 1 and 2) or
		categorical matrix parameters (simulation 3). The standard 
		deviations are given within the parentheses. 
		Both, mean and standard deviation, are with respect to repeat
		simulations. }  
	\centering
	\begin{tabular}{cccccccccc}
		\\\hline\hline
		&\multicolumn{2}{c}{DPM}&
		&\multicolumn{2}{c}{FA} &
		&\multicolumn{2}{c}{DFA}\\
		\cline{2-3} \cline{5-6} \cline{8-9}
		&CMC&MCMC&&CMC&MCMC&&CMC&MCMC \\
		$e_{\Ab}$&
		0.06 (0.03)&0.03 (0.01)&& 0.15 (0.05) & 0.05 (0.06) && 0.08 (0.01)&0.02 (0.00)\\
		$e_\theta$&
		0.00 (0.00)&0.00 (0.00)&& 0.01 (0.01) & 0.02 (0.03) &&0.05 (0.04)&0.01 (0.01)\\
		\hline
	\end{tabular}
	\label{tab:ocdfa2}
\end{table}

\subsection{Simulation 2: Feature allocation using the IBP}
Here, we test the performance of CMC for the FA model in Section \ref{sec:ibpfa}. We generate a data matrix with $n= 800$ SNVs and $p = 5$ tumors. We use the same simulation truth as in \cite{xu2015mad}. We assume $K=4$ subclones. The latent binary matrix $\Ab$ is set as follows: $A_{i1}=1$ for $i=1,\dots,100$, $A_{i2}=1$ for $i=1,\dots,250$, $A_{i3}=1$ for $i=1,\dots, 400$ and $A_{i4}=1$ for $i=1,\dots,600$. We draw $(b_j, \ths_{j1},\ldots, \ths_{jK})\sim Dir(0.2,\pib)$ where $\pib$ is a random permutation of $(1,5,6,10)$. We set $p_0=0.01$ and $N_{ij}=50$, and generate $y_{ij}$ from model \eqref{eq:TH}. The same tempering scheme as in \cite{ni2019parallel} is adopted.

Figures \ref{fig:facmc} and
\ref{fig:famcmc} show the bar plots of the posterior mode
$\widehat{K}$ of the number of features across simulations, under the
CMC (panel (c)) and full MCMC (d) implementations. Similarly to DPM
clustering, CMC tends to slightly overestimate the
number of features compared to MCMC.  Let $\mathcal{H}(\cdot,\cdot)$
denote the Hamming distance of two matrices and let $\#(\cdot)$ denote
the total number of elements in a matrix. We 
define a mis-allocation rate for $\Ab$  as the average Hamming
distance between the estimator and the truth:
$e_{\Ab}=\frac{\mathcal{H}(\widehat{\Ab},\Ab)}{\#(\Ab)}$. In the
case where $\widehat{\Ab}$ has more columns than $\Ab$, we remove
the extra columns from $\widehat{\Ab}$. The error rate $\pi_\th$ that is reported in Table \ref{tab:ocdfa2} for the FA model summarizes the MSE in estimating the proportions $\thb_k^\star$.
%Again, the performance of CMC stays almost the same for $\eps \in [0.1,0.5]$.

\subsection{Simulation 3: 
	Double feature allocation using an IBP prior} 
This simulation considers a CMC approximation of posterior inference
in a DFA model with IBP prior. We generate 
the feature allocation matrix $\Ab$ from an IBP($m$) model with $m=1$ and
sample size $n=1000$.
% \bch Discussion and implementation of a simple consensus Monte Carlo
% algorithm \citep{minsker2014scalable,scott2016bayes,ni2018scalable}
% will be given in Section \ref{sec:disc} and Supplementary Material A
% for large sample size. \ech
%
%\mynote{moved sentence about big data to the end of the previous
%section, w/o the references (still in the latex, commented out).}
The resulting matrix $\Ab$ has $K=6$ columns and $n=1000$ rows. Given $K=6$, we set the
feature-specific parameters, $\Cb\in \{-1,0,1\}^{p\times K}$ with
$p=60$. \bch The heatmaps of $\Ab$ and $\Bb$ are shown in 
Figure S1(b)\&(c), in the Supplementary Materials. \ech The observations
$y_{ij}$ 
are then generated from the sampling model \eqref{eq:cat}.

Figures \ref{fig:dfacmc} and
\ref{fig:dfamcmc} show the bar plots of posterior modes $\widehat{K}$
for the number of features, across simulations for CMC (e) and full
MCMC (f) implementations, respectively.
As before, CMC tends to slightly overestimate the number of features
compared to MCMC.  The error rate $\pi_\th$ that is reported in Table
\ref{tab:ocdfa2} for the
DFA model summarizes the error in estimating the matched column
subsets, i.e., the rows of $\Cb$. 
We use the same definition based on the Hamming distance as for
$e_{\Ab}$. 
%The performance of CMC stays virtually the same for $\eps \in [0.1,0.5]$.

We conclude that the proposed CMC algorithm implements a
useful approximation for posterior inference on random subsets under
widely used BNP models, for problems similar to the simulation
scenarios, which were chosen to mimic the main features of the three
motivating examples.

%\begin{table}	
%	\caption{Operating characteristics for DFA simulations. We report the errors in estimating $\Ab$, $\Bb$, $\Cb$ and $K$ from consensus Monte Carlo  (CMC) versus full Markov chain Monte Carlo (MCMC), averaged over 50 repetitions. The standard deviations of errors in $\Ab$, $\Bb$ and $\Cb$  are 0.026 or below. We test the proposed CMC at different threshold values $\eps= 0.25, 0.30, 0.35, 0.40, 0.45, 0.50$. } 
%	\centering
%		\begin{tabular}{ccccccccc}
%			\\\hline\hline
%			&\multicolumn{6}{c}{CMC with $\eps$}&&MCMC\\
%			&0.25 &0.30 &0.35 &0.40 &0.45 &0.50&\\\cline{2-7}\cline{9-9}
%			$e_A$&0.083 &0.082 &0.082 &0.082 &0.082 &0.081 &&0.048\\
%			$e_B$&0.036 &0.039 &0.040 &0.047 &0.043 &0.040 &&0.017\\
%			$e_C$&0.013 &0.017 &0.020 &0.022 &0.020 &0.020 &&0.002\\
%			$p_K$&0.700 &0.720 &0.780 &0.820 &0.880 &0.880 &&0.900\\
%			\hline
%	\end{tabular}
%	\label{tab:ocdfa}
%\end{table}
%

\section{Applications}
\label{sec:app}

\subsection{MNIST image clustering}
\label{sec:mnist}
\begin{figure}
	\centering
	\subfigure[MNIST handwritten digits]
	{\includegraphics[width=.35\textwidth]{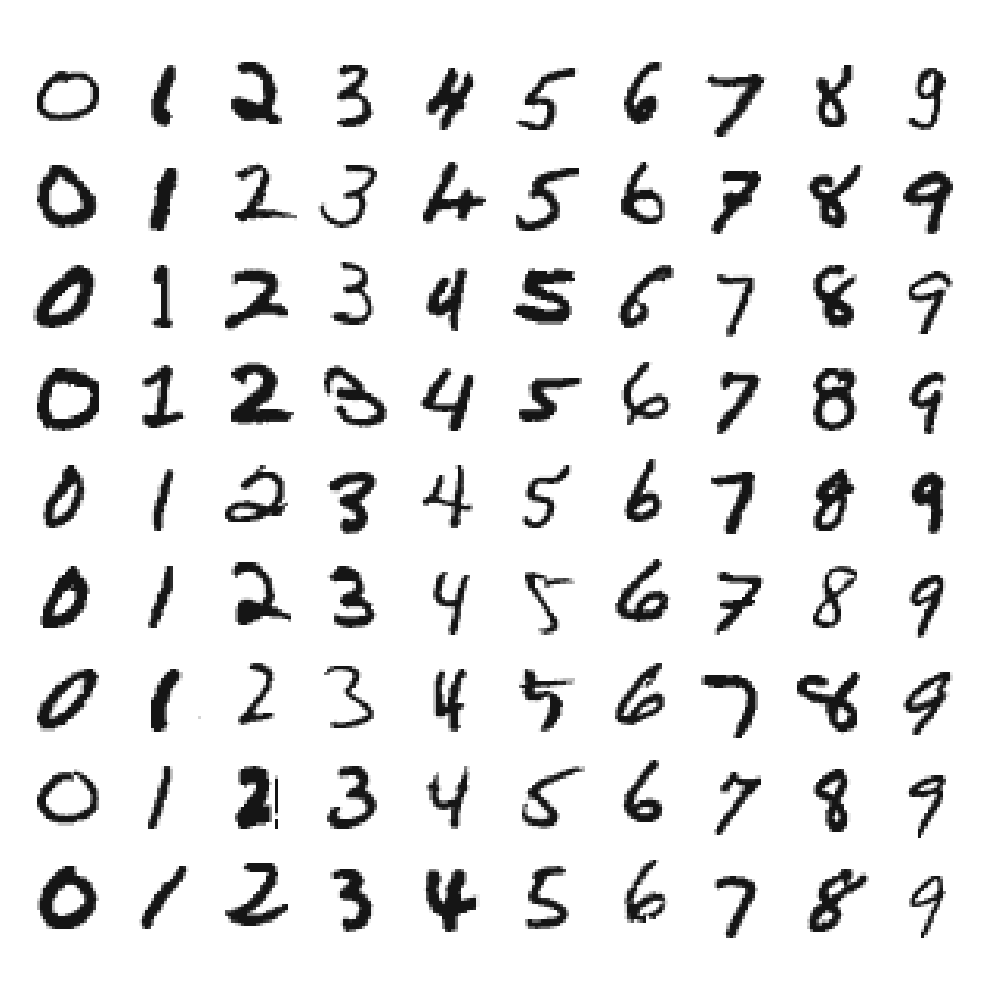}
		\label{fig:dig}}			
	\subfigure[Posterior distribution on $K$]
	{\includegraphics[width=.35\textwidth]{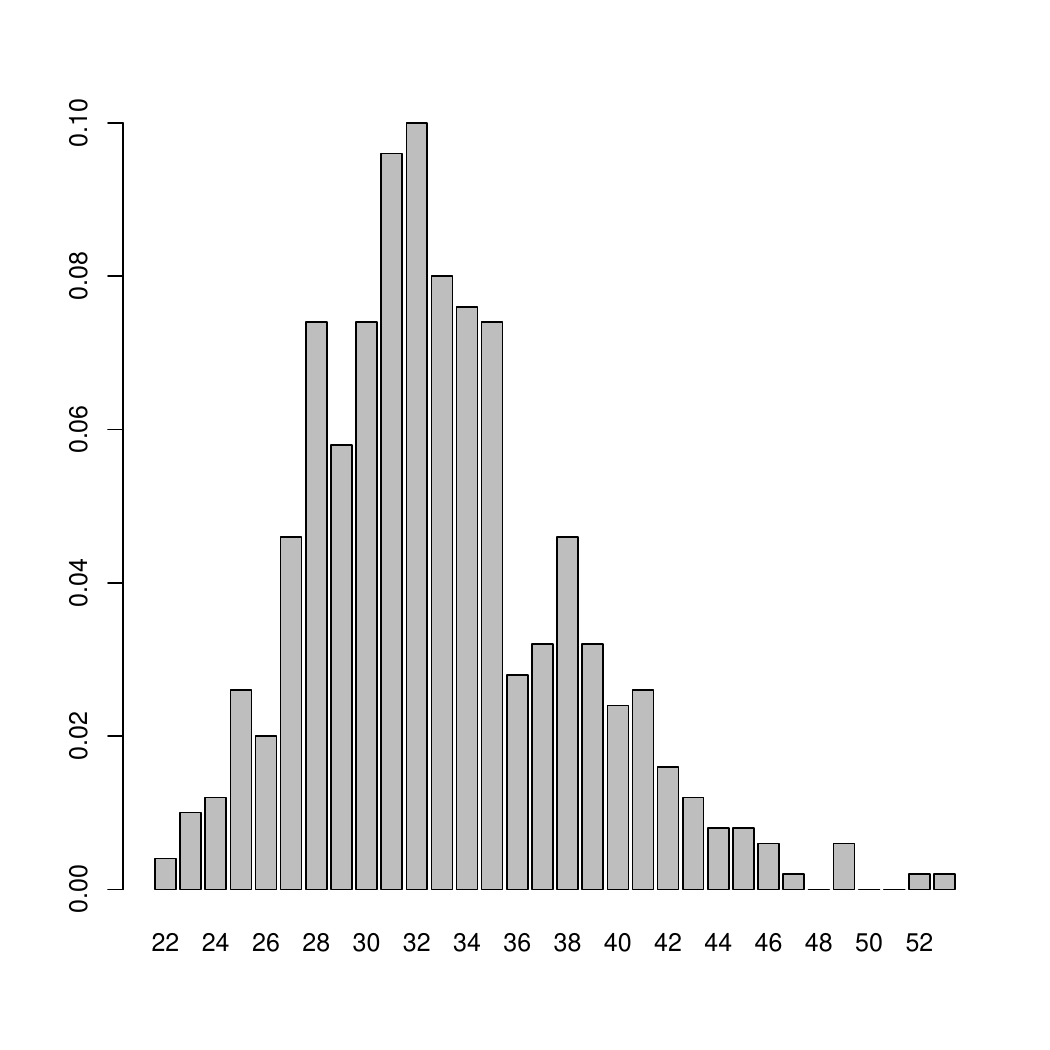}
		\label{fig:pom}}
	\caption{MNIST. (a) A subset of 100 MNIST handwritten
		digits. (b) The posterior distribution of $K$ with
		mode $\widehat{K}=32$.  
	}
	\label{fig:mnist}
\end{figure}

MNIST (\citealt{lecun1998gradient}, \bch \url{http://yann.lecun.com/exdb/mnist/}\ech ) is a 
% well-known image 
dataset of images for classification, containing $n=70,000$
handwritten digits from different writers. 
The data is often used as a benchmark problem for clustering and
classification algorithms. 
Each image has 28$\times$28 pixels that take values
between 0 and 255 representing the grey levels. A subset of the images
are shown in Figures \ref{fig:dig}. We visualize the high-dimensional
image data with a 2D scatter plot \bch 
in Figure S2 (Supplementary Materials) by using \ech the
t-distributed stochastic neighbor embedding (t-SNE,
\citealt{maaten2008visualizing}) with perplexity parameter 75. The
t-SNE algorithm is a non-linear dimension reduction tool that maps
high-dimensional data onto a two- or three-dimensional manifold. 

We randomly split the 70,000 images evenly into 140 shards, each with
500 images, and use one shard as anchor points. We apply the proposed
CMC to the t-SNE transformed data with 5,000 iterations 
(discarding the first 2,500 samples and then thinning out by 5) and $\eps=0.1$. CMC takes less than $10$ minutes to run, whereas
full MCMC simulation takes approximately 30 minutes for only the first 10
iterations. The posterior 
distribution of the number $K$ of clusters is shown in Figure
\ref{fig:pom} which is peaked at a mode at $\widehat{K}=32$.  Conditional on
$\widehat{K}$, the estimated clusters \bch are depicted in 
Figure S2 in the Supplementary Materials. \ech
Keeping in mind that the DP prior favors many singleton clusters,
we drop clusters with fewer than 1\% of the total number of images (9
posterior estimated clusters were singletons), leaving 12 practically
relevant clusters, which only slightly
overestimates the desired number of 10 clusters. Next, we
evaluate the clustering performance, relative to the known truth in
this example,
% Since $\widehat{K}\gg 10$, 
by computing the normalized mutual information (NMI) between the
estimated partition and the true partition (induced from the true labels). \bch Let $\Fb=\{F_k\}_{k=1}^K$ and $\Fb'=\{F_{k'}' \}_{k'=1}^{K'}$ denote two partitions of $[n]$ and let $|F|$ denote the cardinality of a set $F$. NMI is defined as
$\frac{2 \times I(\Fb; \Fb')}{H(\Fb)+H(\Fb')}$, where $I(\cdot,\cdot)=\sum_{k}\sum_{k'}\frac{|F_k\cap F_{k'}'|}{n}\log\frac{n|F_k\cap F_{k'}'|}{|F_k||F_{k'}'|}$
denotes the mutual information between two partitions $\Fb$ and
$\Fb'$ and $H(\Fb)=-\sum_k\frac{|F_k|}{n}\log\frac{|F_k|}{n}$ is the entropy.  \ech NMI is between 0 and 1 with 1
being perfect match between two clusterings. NMI for CMC is 0.76
better than that of K-means 0.72 with $K=32$. Note that the purpose of this application using MNIST data is not to train a classifier or supervised model for the prediction of the 10 digits. Instead, we use the MNIST to examine the feasibility and performance of the proposed CMC algorithm for the clustering (i.e., unsupervised learning) of the relatively large dataset.

To assess the level of approximation of CMC, we use the diagnostic
proposed earlier. Specifically, we sample two
shards and then run CMC (using the same anchor points as before)
as well as full MCMC posterior simulation on the merged dataset of the
two selected shards and anchor points. Repeating the same procedure 69
times, we find an average NMI (between the estimated partitions from
CMC and MCMC) of 0.85 with a standard deviation 0.02, which suggests a
good approximation.  \bch We use NMI rather than more intuitive
measures such as misclassification rate
 to account for the different numbers of clusters under CMC and
  MCMC. 
As a reference, the  NMI is approximately 0.85 when we permute about 7\% cluster assignments of an
estimated partition from MCMC. \ech 

\bch \underline{Sensitivity to $\epsilon$.} \ech To assess the sensitivity of the choice of $\eps$, we repeat the same diagnostic procedure for $\eps=0.05$ and $\epsilon=0.15$. The average NMI is 0.84 for both $\eps$'s.

\bch \underline{Sensitivity to data split.} We repeat the same analysis five times, each time with a different random split of the data. We find the results are stable, e.g. the standard deviation of the NMI (between the
estimated partition and the true partition) is $<0.02$. \ech

%\subsection{Galaxy morphology}
%The full MCMC cannot finish 10 iterations in one day.

\subsection{Inter-tumor Tumor heterogeneity}
\label{sec:itth}
Tumors are genetically heterogeneous, often containing diverse
subclones characterized by genotypic differences. Next-generation
sequencing of tumor samples generates short reads from the genomes of
multiple cells. Since the sequencing is performed in bulk, it is
challenging to reconstruct subclones based on aggregated variant
counts $y_{ij}$ (recall the notation from model \eqref{eq:TH}).
FA models have been proposed in the
literature to infer tumor heterogeneity, including Bayesian
approaches in \cite{lee2015} and \cite{ni2019parallel}. 
Due to the computational limitation of MCMC simulation,
these methods are restricted to a relatively small number of variants
(typically, $n<500$).
\cite{xu2015mad} proposed
a scalable optimization-based algorithm (MAD-Bayes algorithm) to find
a posterior mode. We will compare the proposed CMC with MAD-Bayes. We
use the same pancreatic ductal adenocarcinoma (PDAC) mutation data
analyzed in \cite{xu2015mad}. The PDAC data record the total read
counts $N_{ij}$ and variant read counts $y_{ij}$ at $n=6,599$ SNVs
from $p= 5$ tumors.

\begin{figure}
	\centering
	\includegraphics[width=.4\textwidth]{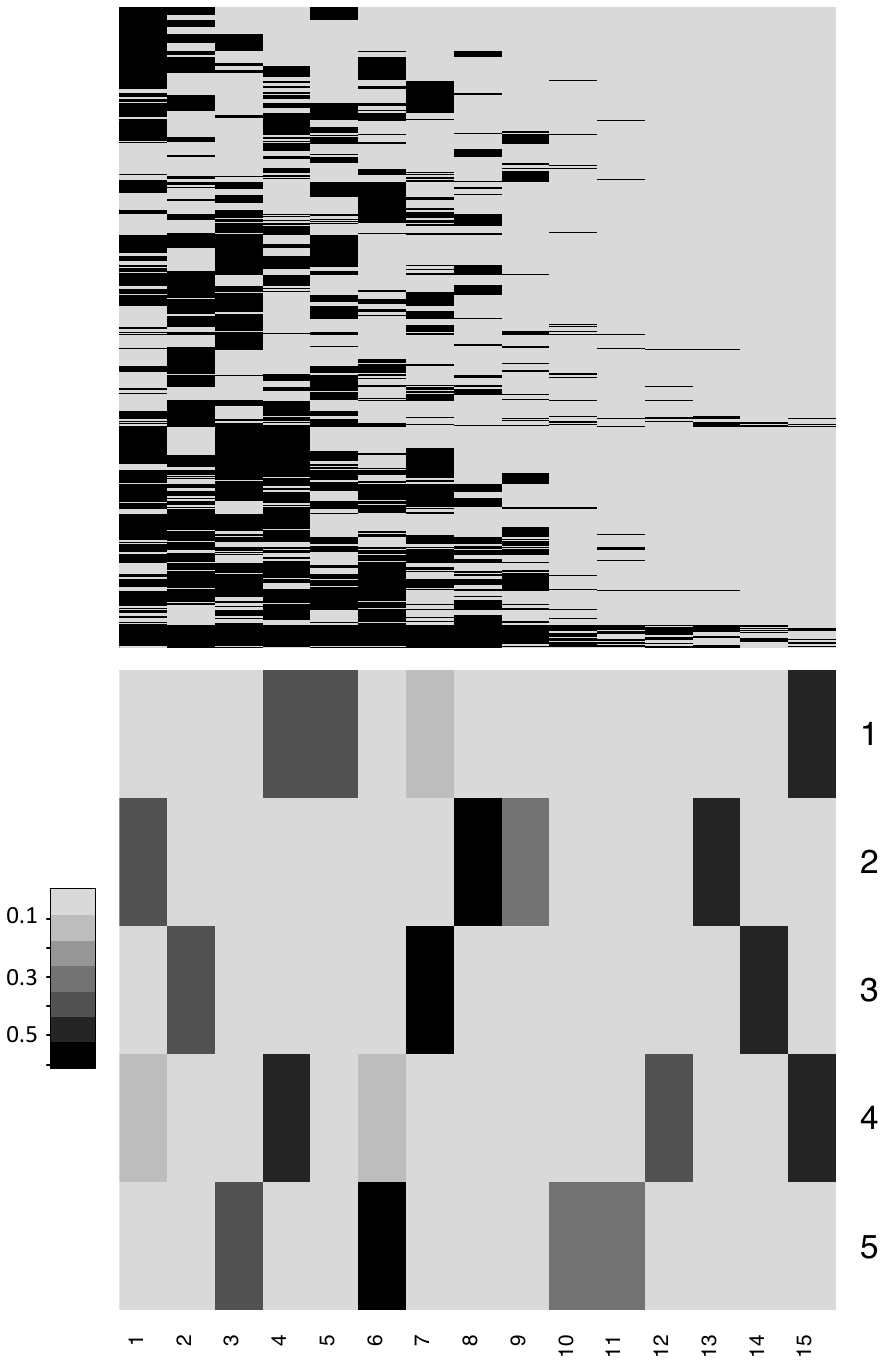}
	\caption{Tumor heterogeneity. The top part of
		the heatmap shows the estimated subclonal genotypes $\Ab$ of the
		selected SNV with dark and light cells representing 1 (mutant) and
		0 (wildtype), 
		respectively. The columns are subclones and the rows are SNVs. The
		bottom part of the heatmap shows the estimated proportions
		$\theta_{tk}^{\star}$ of subclones in each tumor. The columns are
		subclones and the rows are tumors.}  
	\label{fig:hmfa}
\end{figure}

We randomly split the 6,599 SNV's into 33 shards. The first 32
shards have 200 SNVs each and the last shard has 199 SNVs which are
used as anchor points. We apply the proposed CMC with 5,000 iterations
(discarding the first 2,500 as burn-in and then thinning by 5) and the same $\eps=0.1$ as before. 
\bch CMC takes approximately 90 
minutes, whereas the full MCMC is infeasible. \ech We find
15 major subclones across tumors after removing  latent features with
fewer than 5\% SNVs. The estimated feature allocation matrix $\widehat{\Ab}$
is shown as a heatmap at the top of Figure \ref{fig:hmfa}. The
estimated subclone proportions $\theta_{tk}^{\star}$  in each tumor
are shown as a heatmap at the bottom of Figure \ref{fig:hmfa}. The
number of subclones and the ``checkerboard" pattern of the proportions
suggest strong inter-tumor heterogeneity in this data. 

We compare the computational efficiency with MAD-Bayes, an
optimization-based approach. Since each run of MAD-Bayes algorithm may
return different outputs, it is recommended to run the algorithm
repeatedly
\citep{xu2015mad} (e.g. 1,000 times). MAD-Bayes has a regularization parameter
$\lambda^2$ that controls the number $K$ of subclones. The parameter
$\lambda^2$ needs to be carefully tuned over a range of values (say,
30 values). Using the same computer resources as for CMC (i.e. 32
computing cores), \bch it takes more than one day to finish. \ech

\subsection{Electronic health records phenotyping}
\label{sec:ehr}
We consider a large EHR dataset with $n=100,000$ patients from China. The
data are from a physical exam of Chinese residents in 2016. We
extract the blood test results measured on $p=39$ testing items (shown
in Table \ref{tab:names}) and diagnostic codes for diabetes from the
EHR.  We implement inference under a DFA prior and sampling model \eqref{eq:cat}. Each latent feature can be interpreted as a latent disease
that favors symptoms $R_k=\{j \mid C_{jk}=1\}$ and is
related to a subset of patients $F_k=\{i \mid A_{ik}=1\}$. That is,
$\Ab$ describes the patient-disease relationships and $\Cb$ describes
symptom-disease relationships. We follow the same procedure in
\cite{Ni:19} to preprocess the data. Using the reference
range for each test item, we discretize the data and define a symptom
if the value of an item falls beyond the reference range. We fix the
first column of $\Ab$ in the DFA model according to the diabetes
diagnosis. Moreover, since diabetes is clinically associated with high
glucose level, we incorporate this prior information by fixing the
corresponding entry in the first column of $\Cb$. Additional prior
knowledge regarding the symptom-disease relationships is
incorporated. Creatinine and blood urea nitrogen (BUN) are two
important indicators of kidney disease. High levels of these two items
suggest impaired kidney function. We fix the two entries
(corresponding to creatinine and BUN) of the second column of $\Cb$ to
1 and the rest to 0. Similarly, elevated systolic blood pressure and
diastolic blood pressure indicate hypertension, and abnormal levels of
total bilirubin (TB), aspartate aminotransferase (AST) and alanine
aminotransferase (ALT) are indicators of liver diseases. We fix the
corresponding entries of the third and fourth column of $\Cb$.

\begin{figure}
	\centering
	\includegraphics[width=\textwidth]{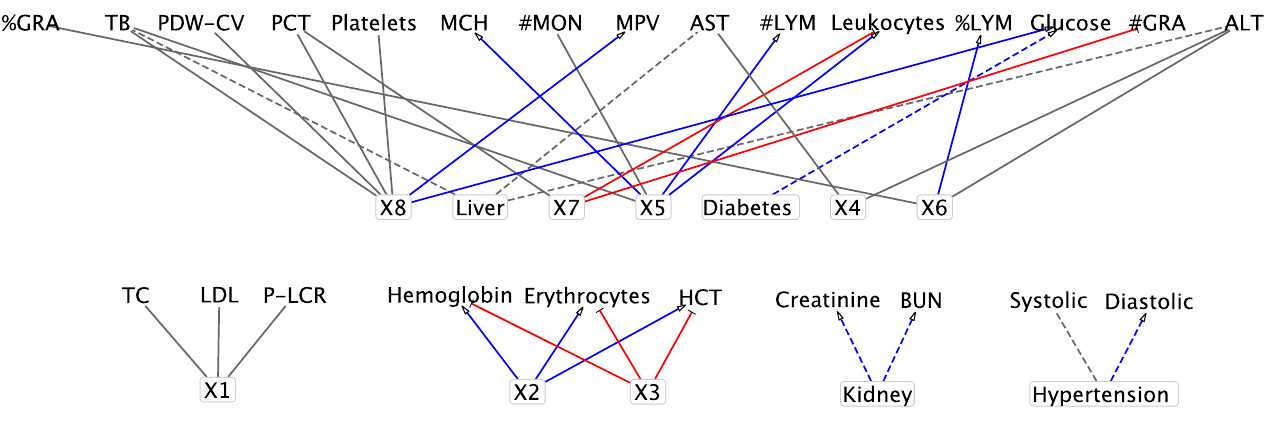}
	\caption{EHR. Bipartite network for symptom-disease relationships from DFA. The diseases are represented by rectangles. Latent diseases are represented by X1$,\dots,$X8. Solid lines are symptom-disease relationships inferred from the data whereas dashed lines are fixed by prior knowledge. Undirected edges indicate the symptoms are binary. Directed edges with bars (arrowheads) indicate the disease causes the symptom to be lower (higher) than normal range.}
	\label{fig:sdnet}
\end{figure}

We randomly split the 100,000 patients evenly into 500 shards, each
with 200 patients, and use one shard as anchor points. We apply the
proposed CMC with 50,000 iterations (discarding the first 25,000 as
burn-in and then thinning out to every 10th) and the same $\eps=0.1$ as before. CMC simulation takes
approximately 1 hour, whereas full MCMC takes 90 minutes for the first
10 iterations.  We find 8 major latent diseases in
addition to the 4 known diseases after removing tiny latent features
(with $<$1\% patients). The estimated symptom-disease relationships
$\widehat{\Cb}$ are represented by a bipartite network in Figure
\ref{fig:sdnet} and also as a heatmap \bch in 
Figure S3 (Supplementary Materials). \ech The estimated patient-disease relationships
$\widehat{\Ab}$ for 1000 randomly selected patients are shown as a
heatmap \bch in Figure S3. \ech

Unlike MNIST or the application to tumor heterogeneity, there is 
no ground truth or alternative implementation for posterior inference
for the full data.
Instead we compare the results with previous results by
\cite{Ni:19} who used a full MCMC implementation for a subset
of 1000 patients from the same dataset. Some of our findings are
consistent with the earlier results, which suggests a good approximation of the
proposed CMC to full MCMC. Moreover, with a hundred times more
observations, we are able to find more interpretable results compared
to theirs.

Latent disease X1 (previously referred to as lipid disorder) is
primarily associated with elevated total cholesterol (TC) and low
density lipoprotein (LDL). Patients with high levels of TC and LDL
have higher risks of heart disease and stroke. Latent diseases X2 and
X3 are associated with the same set of symptoms but with opposite
signs. This interesting result is also found in \cite{Ni:19}
where X2 and X3 were identified as polycythemia and anemia,
respectively. Each of X5, X6 and X8 also finds good correspondence in
\cite{Ni:19} as bacterial infection, viral infection and
thrombocytopenia.

One prevelant latent disease reported in \cite{Ni:19} does
not have a clear interpretation due to the excessive number of
symptoms. They suspect a subset of symptoms like decreased
plateletcrit (PCT), leukocytes and GRA are due to weak immune system
of the elderly population. However, other symptoms are not related to
immune system. With a much larger sample size, we are able to single
out those symptoms without spurious links through the latent disease
X7.

\begin{table}	
	\caption{Blood test items. Acronyms are given within
		parentheses. CV indicates coefficient of variation,
		``dist'' is distribution, ``mn'' is mean,
		``ct''is count, and ``conc" is concentration.} 
	\centering
	\scalebox{.8}{
		\begin{tabular}{lll}
			\\\hline\hline
			alanine aminotransferase (ALT)& 
			aspartate aminotransferase (AST)&
			total bilirubin (TB)\\ 
			total cholesterol (TC)& 
			triglycerides& 
			low density lipoproteins (LDL)\\ 
			high density lipoproteins (HDL)& 
			urine pH (UrinePH)& 
			mn corpuscular hemoglobin conc (MCHC)\\
			\% of monocytes (\%MON) &
			alpha fetoprotein (AFP)& 
			carcinoembryonic antigen (CEA)\\ 
			number of monocytes (\#MON)&
			plateletcrit (PCT)& 
			CV of platelet dist. (PDW-CV) \\  %width coefficient of variation (PDW-CV)\\ 
			\% of eosinophil (\%Eosinophil)&
			basophil ct (\#Basophil)& 
			\% of basophil (\%Basophil)\\
			platelet large cell ratio (P-LCR)& 
			platelets& 
			systolic blood pressure (Systolic)\\ 
			\% of granulocyte (\%GRA)& 
			body temperature (BodyTemperature)&
			{leukocytes}\\
			{hemoglobin}& 
			{creatinine}& 
			{blood urea nitrogen (BUN)}\\ 
			{glucose}& 
			{diastolic blood pressure (Diastolic)}& 
			{heart rate (HeartRate)}\\
			{erythrocytes}& 
			{hematocrit (HCT)}& 
			{uric acid (UricAcid)}\\ 
			{\% of lymphocyte (\%LYM)}&
			{mn corpuscular volume (MCV)}& 
			mn corpuscular hemoglobin (MCH)\\ 
			{lymphocyte ct (\#LYM)}& 
			{granulocyte ct (\#GRA)}&  
			{mn platelet volume (MPV)} 
	\end{tabular}}
	\label{tab:names}
\end{table}

\section{Discussion}
\label{sec:disc}
We have developed a simple CMC algorithm for fast approximate
posterior inference of BNP models. The proposed CMC is a general
algorithm in the sense that it can be used to scale up practically any
Bayesian clustering and feature allocation methods. CMC runs MCMC on
subsets of observations in parallel and aggregates the Monte Carlo
samples. The main idea of this paper is using a subset of observations
as anchor points to merge clusters or latent features from different
machines. The aggregation step has a tuning parameter $\eps$ which is
not influential of the results in our simulations and applications.  \bch 
We have focused on point estimation throughout the paper. But the
proposed algorithm provides an efficient approximation of the
 entire posterior distribution. The output of the proposed
  algorithm is a consensus Monte Carlo sample for the whole data posterior,
  which can be used for any Monte Carlo based posterior inference. 
For example, to find the 90\% credible interval of a
continuous parameter, one can take the  5\% and 95\% sample quantiles
of the consensus Monte Carlo sample. For discrete structures such as
partitions, one can use the approach in \cite{wade2018bayesian} to
construct credible balls of partitions, again using the consensus Monte
Carlo sample. 

The anchor points are randomly chosen. Overall, we find 
 that inference remains robust with respect to the choice of the
  anchor points, as long as the number of anchor points is sufficiently large. The
analysis in Section \ref{sec:mnist} leads to similar results (not
shown) under different random splits of the data. 
 Alternatively to using randomly chosen anchor points, one may use
big-data summarization approaches to guide the choice, using for
example the notion of a ``coreset'' developed in 
\cite{huggins2016coresets,campbell2019automated}. 
Alternatively, \cite{mak2018support} proposed a small set of
``support points" to represent a continuous probability
distribution.  \ech

% We will leave the exploration of the effectiveness of
% these two approaches in big-data clustering and feature allocation  as
% our future work. \ech 

We explicitly address the problem of large sample size but have not
considered the issue of high-dimensionality.
 % which we plan to deal with in our future work.  
We discussed  the approach for random subsets in random partitions and
(double) feature allocation. In the motivating examples we used DPM 
and IBP priors. But the algorithm remains equally valid for any other
prior model. The approach with common anchors remains useful also for any other
models that involve random subsets that can be split and merged in a
similar fashion, including latent trait models, finite
mixture models, and finite feature allocation.

\section*{Acknowledgment}
The authors express thanks to Andres Christen (CIMAT, Guanajuato,
Mexico) for first suggesting the use of anchor points for CMC with
random subsets and to David Jones (Texas A\&M) for useful discussion on case studies.  PM and YJ are partly supported by NIH R01 CA132897.

\section*{Supplementary Materials}

\begin{description}
	
	\item [Additional figures] S1 through S3 for simulations and applications. (.pdf file)
	
	\item[Software] to merge Monte Carlo samples of clusters and latent features and to generate all simulated data.

\end{description}

\bibliographystyle{apalike}
\bibliography{bib_bigdfa}
\end{document}